\documentclass{aa} 

\usepackage{graphicx}
\usepackage{txfonts}

\usepackage{natbib}
\usepackage{dcolumn}
\usepackage{amssymb}

\newcolumntype{d}{D{.}{.}{-1}}

\begin{document} 

  \title{Asteroid models from the Lowell Photometric Database}

  \author{J. \v{D}urech         \inst{1}        \and
          J. Hanu\v{s}          \inst{2,3}      \and
          D. Oszkiewicz         \inst{4}        \and
          R. Van\v{c}o          \inst{5}        
          }

  \institute{Astronomical Institute, Faculty of Mathematics and Physics, Charles University in Prague, V Hole\v{s}ovi\v{c}k\'ach 2, 180\,00 Prague 8, Czech Republic\\
             \email{durech@sirrah.troja.mff.cuni.cz}
        \and
             Centre National d'\'Etudes Spatiales, 2 place Maurice Quentin, 75039 Paris cedex 01, France
        \and
             Laboratoire Lagrange, UMR7293, Universit\' e de la C\^ ote d'Azur, CNRS, Observatoire de la C\^ ote d'Azur, Blvd de l'Observatoire, CS 34229, 06304 Nice cedex 04, France
        \and
             Astronomical Observatory Institute, Faculty of Physics, A. Mickiewicz University, S{\l}oneczna 36, 60-286 Pozna{\'n}, Poland
        \and
             Czech National Team
             }

  \date{Received ?; accepted ?}

  \abstract
  {Information about shapes and spin states of individual asteroids is important for the study of the whole asteroid population. For asteroids from the main belt, most of the shape models available now have been reconstructed from disk-integrated photometry by the lightcurve inversion method.}
  {We want to significantly enlarge the current sample ($\sim$\,350) of available asteroid models.}
  {We use the lightcurve inversion method to derive new shape models and spin states of asteroids from the sparse-in-time photometry compiled in the Lowell Photometric Database. To speed up the time-consuming process of scanning the period parameter space through the use of convex shape models, we use the distributed computing project Asteroids@home, running on the Berkeley Open Infrastructure for Network Computing (BOINC) platform. This way, the period-search interval is divided into hundreds of smaller intervals. These intervals are scanned separately by different volunteers and then joined together. We also use  an alternative, faster, approach when searching the best-fit period by using a model of triaxial ellipsoid. By this, we can independently confirm periods found with convex models and also find rotation periods for some of those asteroids for which the convex-model approach gives too many solutions.}
  {From the analysis of Lowell photometric data of the first 100,000 numbered asteroids, we derived 328 new models. This almost doubles the number of available models. We tested the reliability of our results by comparing models that were derived from purely Lowell data with those based on dense lightcurves, and we found that the rate of false-positive solutions is very low. We also present updated plots of the distribution of spin obliquities and pole ecliptic longitudes that confirm previous findings about a non-uniform distribution of spin axes. However, the models reconstructed from noisy sparse data are heavily biased towards more elongated bodies with high lightcurve amplitudes.}
  {The Lowell Photometric Database is a rich and reliable source of information about the spin states of asteroids. We expect  hundreds of other asteroid models for asteroids with numbers larger than 100,000 to be derivable from this data set. More models will be able to be reconstructed when Lowell data are merged with other photometry.}


  \maketitle

  \section{Introduction}

  Large all-sky surveys like Catalina, Pan-STARRS, etc. image the sky every night to discover new asteroids and detect those that are potentially hazardous. The main output of these surveys is a steadily increasing number of asteroids with known orbits. Apart from astrometry that is used for orbit computation, these surveys also produce photometry of asteroids. This photometry contains, in principle, information about asteroid rotation, shape, and surface properties. However, because of its poor quality (when compared with a dedicated photometric measurements of a single asteroid) the signal corresponding to asteroid's rotation is usually drowned in noise and systematic errors. However, there have been recent attempts to use sparse-in-time photometry to reconstruct the shape of asteroids. \cite{Kaa:04} has shown that sparse photometry can be used to solve the lightcurve inversion problem and further simulations confirm this \citep{Dur.ea:05, Dur.ea:07}. Afterwards, real sparse data were used either alone or in combination with dense lightcurves and new asteroid models were derived \citep{Dur.ea:09, Cel.ea:09, Han.ea:11, Han.ea:13b}. The aim of these efforts was to derive new unique models of asteroids, i.e., their sidereal rotation periods, shapes, and direction of spin axis. 
  
  Another approach to utilize sparse data was to look for changes in the mean brightness as a function of the aspect angle, which led to estimations of spin-axis longitudes for more than 350,000 asteroids \citep{Bow.ea:14} from the so-called Lowell Observatory photometric database \citep{Osz.ea:11}. 
  
  In this paper, we show that the Lowell photometric data set can be also be used for solving the full inversion problem. By processing Lowell photometry for the first 100,000 numbered asteroids, we derived new shapes and spin states for 328 asteroids, which almost doubles the number of asteroids for which the photometry-based physical model is known.   
  
  We describe the data, the inversion method, and the reliability tests in Sect.~\ref{sec:method}, the results in Sect.~\ref{sec:results}, and we conclude in Sect.~\ref{sec:conclusion}

  \section{Method}
  \label{sec:method}

  The lightcurve inversion method of \cite{Kaa.ea:01} that we applied was reviewed by \cite{Kaa.ea:02c} and more recently by \cite{Dur.ea:15b}. We used the same implementation of the method as \cite{Han.ea:11}, where the reader is referred to for details. Here we describe only the general approach and the details specific for our work.
  
  \subsection{Data}
  
  As the data source, we used the Lowell Observatory photometric database \citep{Bow.ea:14}. This is photometry provided to Minor Planet Centre (MPC) by 11 of the  largest surveys that were re-calibrated in the V-band using the accurate photometry of the Sloan Digital Sky Survey. Details about the data reduction and calibration can be found in \cite{Osz.ea:11}. The data are available for about $\sim 330,000$ asteroids. Typically, there are several hundreds of photometric points for each asteroid. The length of the observing interval is $\sim 10$--15 years. The largest amount of data is for the low-numbered asteroids and decreases with increasing asteroid numbers. For example, the average number of data points is $\sim 480$ for asteroids with number $< 10,000$ and $\sim 45$ for those $> 300,000$. The accuracy of the data is around 0.10--0.20\,mag.

  For each asteroid and epoch of observation, we computed the asteroid-centric vectors towards the Sun and the Earth in the Cartesian ecliptic coordinate frame -- these were needed to compute the illumination and viewing geometry in the inversion code.
  
  \subsection{Convex models}

  To derive asteroid models from the optical data, we used the lightcurve inversion method of \cite{Kaa.Tor:01} and \cite{Kaa.ea:01}, the same way as \cite{Han.ea:11}. Essentially, we searched for the best-fit model by densely scanning the rotation period parameter space. We decided to search in the interval of 2--100 hours. The lower limit roughly corresponds to the observed rotation limit of asteroids larger than $\sim 150$\,m \citep{Pra.ea:02b}, the upper limit was set arbitrarily to cover most of the rotation periods for asteroids determined so far. For each trial period, we started with ten initial pole directions that were isotropically distributed on a sphere. This turned out to be enough  not to miss any local minimum in the pole parameter space. In each period run, we recorded the period and $\chi^2$ value that correspond to the best fit. Then we looked for the global minimum  of $\chi^2$ on the whole period interval and tested the uniqueness and stability of this globally best solution (see details in Sect.~\ref{sec:tests}).

  For a typical data set, the number of trial periods is 200,000--300,000, which takes about a month on one CPU. Because the number of asteroids we wanted to process was $\sim 100,000$, the only way  to finish the computations in a reasonable time was to use tens of thousands of CPUs. For this task, we used the distributed computing project Asteroids@home.\footnote{\url{http://asteroidsathome.net}}
  
  \subsection{Asteroids@home}

  Asteroids@home is a volunteer-based computing project built on the Berkeley Open Infrastructure for Network Computing (BOINC) platform. Because the scanning of the period parameter space is the so-called embarrassingly parallel problem, we divided the whole interval of 2--100 hours into smaller intervals (typically hundreds), which  were searched individually on the computers of volunteers connected to the project. The units sent to volunteers had about the same CPU-time demand. Results from volunteers were sent back to the BOINC server and validated. When all units belonging to one particular asteroid were ready, they were connected and the global minimum was found. The technical details of the project are described in \cite{Dur.ea:15}

  \subsection{Ellipsoids}

  To find the rotation period in sparse data, we also used  an alternative approach that was based on the triaxial ellipsoid shape model and a geometrical light-scattering model \citep{Kaa.Dur:07}. Its advantage is that it is much faster than using convex shapes because the brightness can be computed analytically \citep[it is proportional to the illuminated projected area,][]{Ost.Con:84}. On top of that, contrary to the convex modelling, all shape models automatically fulfill the physical condition of rotating along the principal axis with the largest momentum of inertia. The accuracy of this simplified model is sufficient to reveal the correct rotation period as a significant minimum of $\chi^2$ in the period parameter space. That period is then used as a start point for the convex inversion for the final model. In many cases when the convex models gives many equally good solutions with different periods, this method provides a unique and correct rotation period.

  \subsection{Restricted period interval}

  As mentioned above, the interval for period search was 2--100 hours. However, for many asteroids, their rotation period is known from observations of their lightcurves. The largest database of asteroid rotation periods is the Lightcurve Asteroid Database (LCDB) compiled by \cite{War.ea:09} and regularly updated at \url{http://www.minorplanet.info/lightcurvedatabase.html}. If we take information about the rotation period as an a priori constraint, we can narrow the interval of possible periods and significantly shrink the parameter space. For this purpose, we used only reliable period determinations from LCDB with quality codes U equal to 3, 3-, or 2+. However, even for these quality codes, the LCDB period can be wrong \citep[for examples see][]{Mar.ea:15} resulting in a wrong shape model. For quality codes 3 and 3-, we restricted the search interval to $P \pm 0.05 P$, where $P$ was the rotation period reported in LCDB. Similarly for U equal to 2+, we restricted the search interval to $P \pm 0.1 P$. We applied this approach to both convex- and ellipsoid-based period search.

  \subsection{Tests}
  \label{sec:tests}

  For each periodogram, there is formally one best model that corresponds to the period with the lowest $\chi^2$. However, the global minimum in $\chi^2$ has to be significantly deeper than all other local minima to be considered as a reliable solution, rather than just a random fluctuation. We could not use formal statistical tools to decide whether the lowest $\chi^2$ value is statistically significant or not, because the data were also affected  by systematic errors. Instead, to select only robust models, we set up several tests, which  each model had to pass to be considered as a reliable model.

  \begin{enumerate}
   \item The lowest $\chi^2$ corresponding to the rotation period $P_\text{min}$ is at least 5\% lower than all other $\chi^2$ values for periods outside the $P_\text{min} \pm 0.5 P_\text{min}^2 / \Delta T$ interval, where $\Delta T$ is the time span of observations \citep{Kaa:04}. The value of 5\% was chosen such that the number of unique models was as large as possible while keeping the number of a false positive solution very low ($\sim 1$\%). The comparison was done with respect to models in DAMIT (see Sect.~\ref{sec:models_comparison}).
   \item When using convex models for scanning the period parameter space, we ran the period search for two resolutions of the convex model -- the degree and order of the harmonics series expansion that parametrized the shape was three or six. The periods $P_\text{min}$ corresponding to these two resolutions had to agree within their errors (and both had to pass the test nr.~1).
   \item Because we realized that $P_\text{min} \gtrsim 20$\,h often produced false positive solutions, we accepted only models with $P_\text{min}$ shorter than 20 hours (when there was no information about the rotation period from LCDB). 
   \item For a given $P_\text{min}$, there are no more than two distinct (farther than $30^\circ$ apart) pole solutions with $\chi^2$ at least 5\% deeper than other poles.
   \item Because of the geometry limited close to the ecliptic plane, two models that have the same pole latitudes $\beta$ and pole longitudes $\lambda$ that are different by $180^\circ$ provide the same fit to disk-integrated data, and they cannot be distinguished from each other \citep{Kaa.Lam:06}. Therefore we accepted only such solutions that fulfilled the condition that if there were two pole directions $(\lambda_1, \beta_1)$ and $(\lambda_2, \beta_2)$, the difference in ecliptic latitudes $|\beta_1 - \beta_2|$ has to be less than $50^\circ$ and the difference in ecliptic longitudes $\mod (|\lambda_1 - \lambda_2|, 180^\circ)$ has to be larger than $120^\circ$.
   \item The ratio of the moment of inertia along the principal axis to that along the actual rotation axis should be less than 1.1. Otherwise the model is too elongated along the direction of the rotation axis and it is not considered  a realistic shape.
   \item For each asteroid that passed the above test, we created a bootstrapped lightcurve data set by randomly selecting the same number of observations from the original data set. This new data set was processed the same way as the original one (using either convex shapes or ellipsoids for the period search) and the model was considered stable only if the best-fit period $P_\text{min}$ from the bootstrapped data agreed with that from the original data. 
   \item We also visually checked all shape models, periodograms, and fits to the data to be sure that the shape model looked realistic and that there were no clear problems with the data and residuals. In some rare cases we rejected models that formally fitted the data, passed all the test, but were unrealistically elongated or flat.
  \end{enumerate}

  \begin{figure}
   \begin{center}
    \includegraphics[width=\columnwidth]{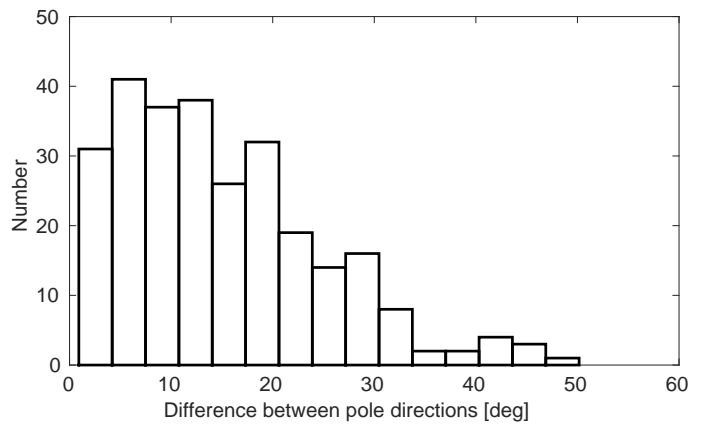}
    \caption{Histogram of differences between the pole directions of models derived from Lowell data and those archived in DAMIT.}
    \label{fig:histogram_poles}
   \end{center}
  \end{figure}

  \section{Results}
  \label{sec:results}

  \subsection{Comparison with independent models}
  \label{sec:models_comparison}

  From all $\sim 600$ models that successfully passed the tests described in Sect.~\ref{sec:tests}, some were already modeled from other photometric data and the models were stored in the Database of Asteroid Models from Inversion Techniques \citep[DAMIT\footnote{\url{http://astro.troja.mff.cuni.cz/projects/asteroids3D}},][]{Dur.ea:10}. For this subset, we could compare our results from an inversion of Lowell data with independent models (assumed to be reliable) from DAMIT. In total, there were 279 models in DAMIT for comparison. For these models, we computed the difference between the DAMIT and Lowell rotation periods and also the difference between the pole directions. Out of this set, almost all (275 models) have the same rotation periods (within the uncertainties) and the pole differences $< 50^\circ$ of arc. The histogram of pole differences between DAMIT and our models is shown in Fig.~\ref{fig:histogram_poles}. Although there are some asteroids for which we got differences as large as $\sim 40$--$50^\circ$, the mean value is $15^\circ$ and the median $13^\circ$, which can be interpreted as a typical error in the pole determination that was based on Lowell data, assuming that the poles from DAMIT have smaller errors (typically 5--10$^\circ$). 

\begin{figure}
   \begin{center}
    \includegraphics[width=\columnwidth]{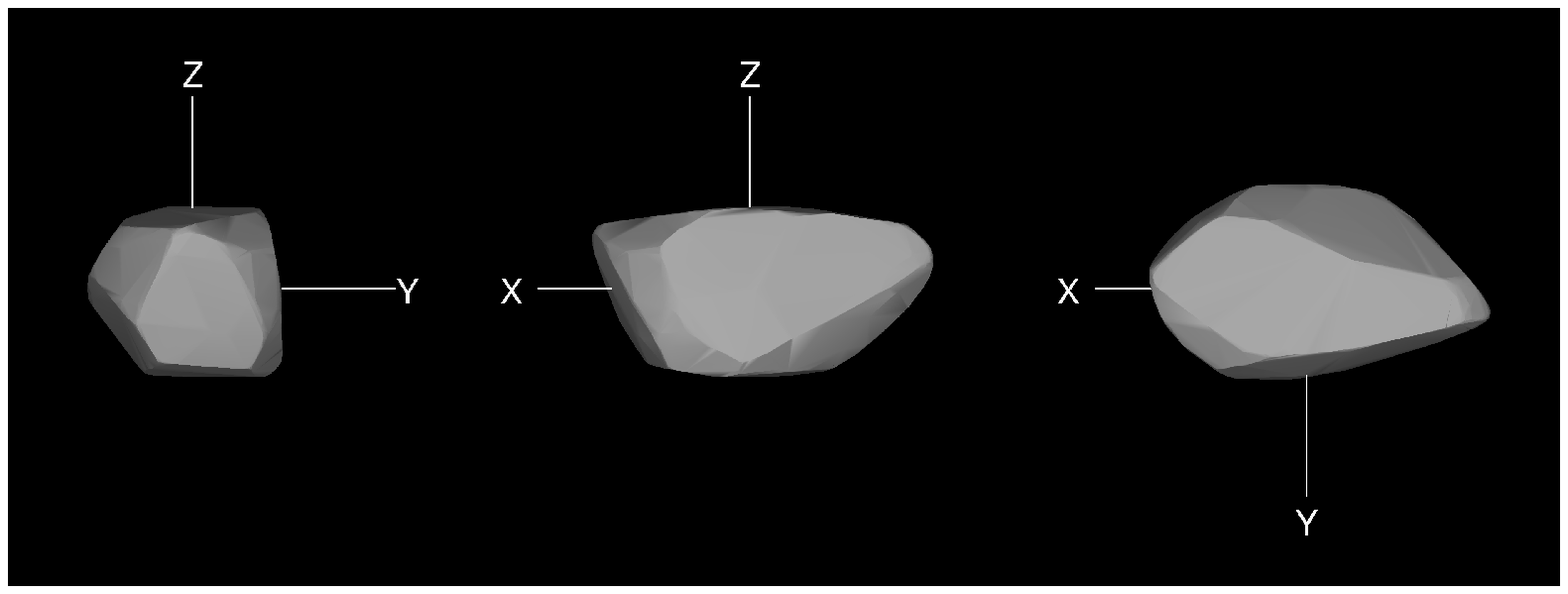}\\
    \includegraphics[width=\columnwidth]{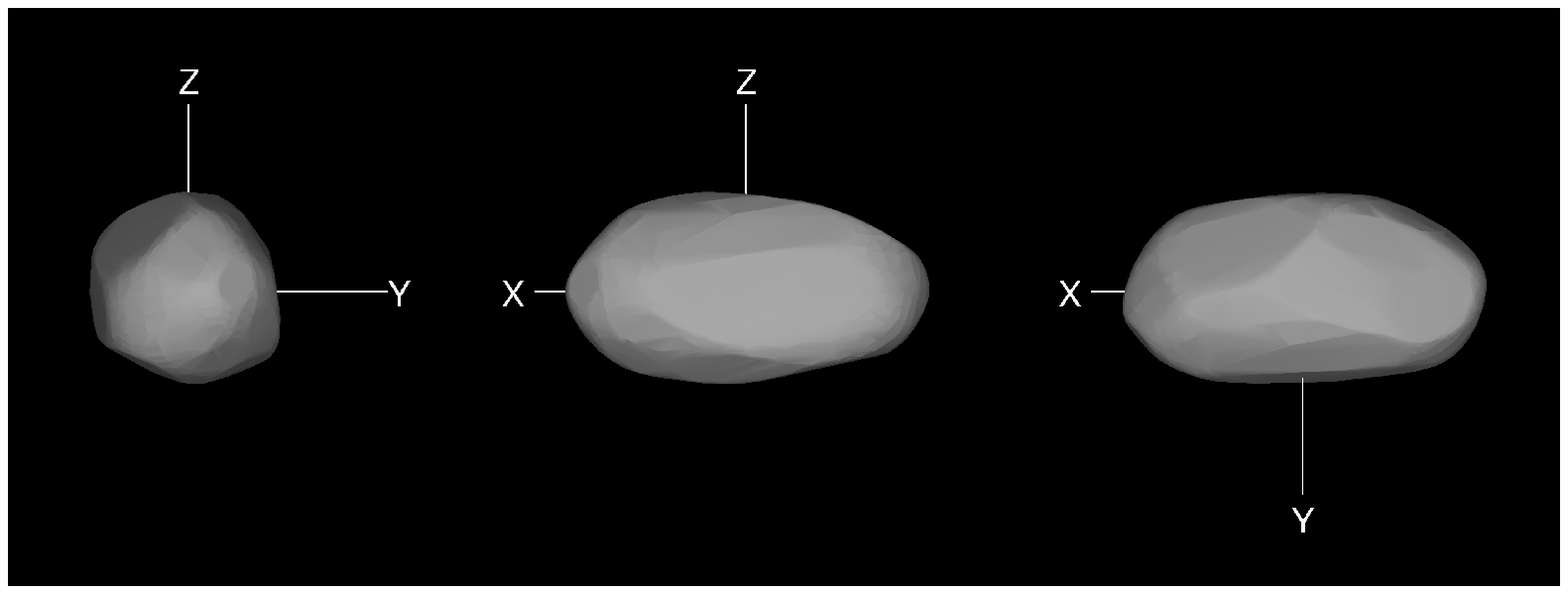}\\
    \caption{Comparison between the shape model of (63)~Ausonia reconstructed from Lowell sparse data (top) and from dense lightcurves (bottom).}
    \label{fig:Ausonia}
   \end{center}
  \end{figure}

  As an example of the difference between shape models, we show results for asteroid (63)~Ausonia. In Fig.~\ref{fig:Ausonia}, we compare our shape model, which we derived from Lowell sparse photometry, with that obtained by inversion of dense lightcurves \citep{Tor.ea:03}. In general, the shapes derived from  sparse photometry are more angular than those derived from dense lightcurves and often have artificial sharp edges. 

  The four asteroids (5)~Astraea, (367)~Amicita, (540)~Rosamunde, and (4954)~Eric, for which we got different solutions to DAMIT, are discussed below. We also discuss the five asteroids -- (1753)~Mieke, (2425)~Shenzen, (6166)~Univsima, (11958)~Galiani, and (12753)~Povenmire -- for which there is no model in DAMIT, but the period we derived from the Lowell data does not agree with the data in LCDB.

  \paragraph{(5) Astraea}
  From Lowell data, we got two pole directions $(\lambda, \beta) = (121^\circ, -20^\circ)$ and $(296^\circ, -15^\circ)$, the former being about $60^\circ$ away from the DAMIT model of \cite{Han.ea:13a} with the pole $(126^\circ, 40^\circ)$. The DAMIT model agrees with the adaptive optics data as well as with the occultation silhouette from 2008 and it is not clear why there is so large a difference in the pole direction, while the rotation periods are the same and the number of Lowell photometric points is also large (447 points).

  \paragraph{(367) Amicita}
  The model derived from Lowell data has two pole solutions $(17^\circ, -52^\circ)$ and $(194^\circ, -45^\circ)$ and rotation period of 5.05578\,h, while the DAMIT model of \cite{Han.ea:11} has prograde rotation with poles of $(21^\circ, 32^\circ)$ and $(203^\circ, 38^\circ),$ with a significantly different period of 5.05502\,h. However, the DAMIT model is based on sparse data from US Naval Observatory and Catalina and only two pieces of lightcurve by \cite{Wis.ea:97} and it might not be correct.

  \paragraph{(540) Rosamunde}
  Although the periodogram obtained with the convex model approach shows a minimum for 9.34780\,h -- the same as the DAMIT model of \cite{Han.ea:13c} -- this minimum was not deep enough to pass the test nr.~1. However, the second-best minimum for a convex model at 7.82166\,h appeared as the best solution for the ellipsoid approach and passed all tests leading to a wrong model.

  \paragraph{(4954) Eric}
  The DAMIT model of \cite{Han.ea:13b} has a pole direction of $(86^\circ, -55^\circ)$, which is almost exactly opposite to our value of $(261^\circ, 70^\circ)$. Moreover, even the rotation periods are different by about 0.0003\,h, which is more than the uncertainty interval. 

  \paragraph{(1753) Mieke}
  The rotation period of 8.9\,h was determined by \cite{Lag:78} from two (1.5 and 5 hours) noisy lightcurves. Given the quality of the data, this period is not in contradiction with our value of 10.19942\,h.

  \paragraph{(2425) Shenzen}
  The rotation period of $14.715 \pm 0.012$\,h was determined by \cite{Haw.Dit:08}. Our value of 9.83818\,h is close to 2/3 of their. In the periodogram, there is no significant minimum around 14.7\,h. 

  \paragraph{(6166) Univsima}
  The lightcurve is published online in the database of R.~Behrend\footnote{\url{http://obswww.unige.ch/~behrend/page5cou.html}}. However, the period of 9.6\,h is based on only 12 points, which  covers about half of the reported period, so we think that this preliminary result is not in contradiction with our period of $\sim 11.4$\,h.

  \paragraph{(11958) Galiani}
  This asteroid was observed by \cite{Cla:14}, who determined the period $9.8013 \pm 0.0023$\,h, which does not agree with our value of 8.24720\,h. The reason is not clear, because the data of \cite{Cla:14} seem to fit this period correctly.  We do not see any significant minimum in $\chi^2$ near 9.8\,h in the periodogram.

  \paragraph{(12753) Povenmire}
  The period of 12.854\,h reported in the LCDB is based on the observations of \cite{Gar:04}. However, according to the same author,\footnote{\url{http://brucegary.net/POVENMIRE/}} the correct rotation period that is based on observations from 2010 is $17.5752 \pm 0.0008$\,h, which agrees with our value.

  In summary, the frequency of false positive solutions that pass all reliability tests seems to be sufficiently low, around a few percent. However, the sample of models in DAMIT that we use for comparison is itself biased against low-amplitude long-period asteroids \citep{Mar.ea:15}, so the real number of false positive solutions might be higher.
  
  \subsection{New models}

  After applying all the tests described in Sect.~\ref{sec:tests}, we selected only those asteroids with no model in DAMIT for publication. These are listed in Tables~\ref{tab:models} (models from full interval 2--100 hours) and \ref{tab:models_interval} (models derived from a restricted period interval). The tables list the pole direction(s) (one or two models), the sidereal rotation period (with uncertainty corresponding to the order of the last decimal place). The C/E code means the method by which $P_\text{min}$ was found -- convex models (C) or ellipsoids (E). In some cases, both methods independently gave  the same value of $P_\text{min}$ (then CE code). All new shape models and the photometric data are available in DAMIT.
  
  For some of these asteroids, \cite{Han.ea:15b} obtained independent models by applying the same lightcurve inversion method on sparse data, which they combined with dense lightcurves. These asteroids (not yet published in DAMIT) are marked by asterisk in the Tables~\ref{tab:models} and \ref{tab:models_interval}. For all of them (56 in total), our rotation periods agree with those of \cite{Han.ea:15b} within their uncertainties and pole directions differ by 10--20 degrees on average. By way of comparison, this is a similar result  to the DAMIT models (Sect.~\ref{sec:models_comparison}, Fig.~\ref{fig:histogram_poles}) and it independently confirms the reliability of our models based on only Lowell data.
  
  \begin{figure}
   \begin{center}
    \includegraphics[width=\columnwidth]{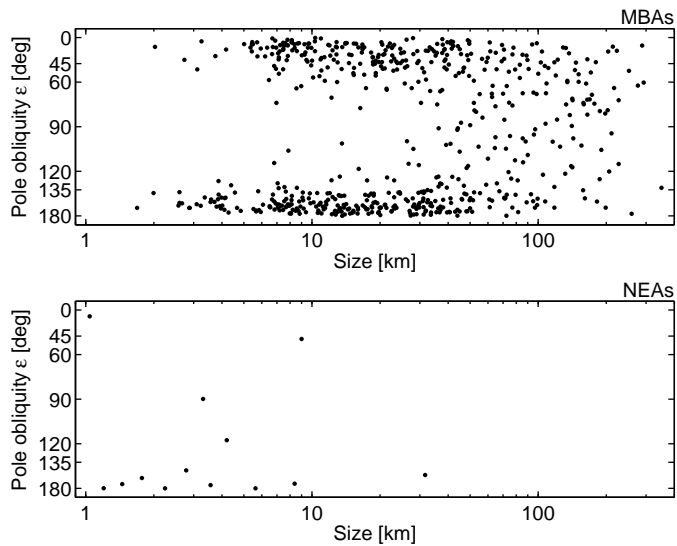}
    \caption{The distribution of pole obliquity $\varepsilon$ as a function of size for 575 main-belt and 13 near-Earth asteroids.}
    \label{fig:obliquity_vs_size}
   \end{center}
  \end{figure}
 
  \begin{figure}
   \begin{center}
    \includegraphics[width=\columnwidth]{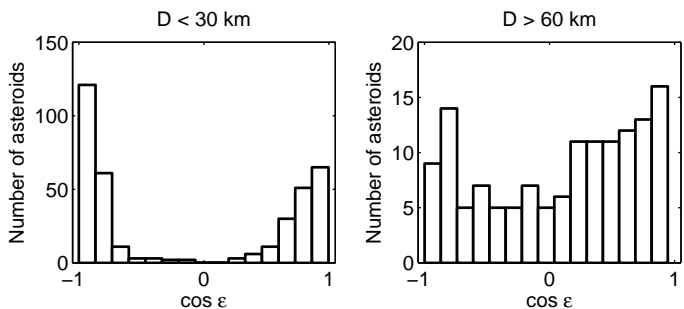}
    \caption{Histograms of the distribution of pole obliquities $\varepsilon$ for asteroids with diameters $< 30$\,km and $> 60$\,km, respectively.}
    \label{fig:histogram_obliquity}
   \end{center}
  \end{figure}

  \subsection{Statistics of pole directions}
  
  Together with models from DAMIT, we now have a sample of shape models for 717 asteroids (685~MBAs, 13~NEAs, 10~Mars-crossers, 7~Hungarias, 1~Hilda, and 1~Trojan). The statistical analysis of the pole distribution confirms the previous findings. Namely, the distribution of spin directions is not isotropic \citep{Kry.ea:07}. Moreover, the distribution of pole obliquities (an angle between the spin vector and the orbital plane) depends on the size of an asteroid. We plot the dependence of obliquity on the size in Fig.~\ref{fig:obliquity_vs_size} for main-belt (MBAs) and near-Earth (NEAs) asteroids. There is a clear trend of smaller asteroids  clustering towards extreme values of obliquity. This was explained by \cite{Han.ea:11} as YORP-induced evolution of spins \citep{Han.ea:13b}. The distribution of obliquities is not symmetric around $90^\circ$ (Fig.~\ref{fig:histogram_obliquity}). As noticed by \cite{Han.ea:13b}, the retrograde rotators are more concentrated to $-90^\circ$, probably because prograde rotators are affected by resonances. For larger asteroids, there is an excess of prograde rotators that might be primordial \citep{Kry.ea:07, Joh.Lac:10}.

  However, the current sample of asteroid models is far from being representative of the whole asteroid population. Because the period search in sparse data is strongly dependent on the lightcurve amplitude -- the larger the amplitude the easier is to detect the correct rotation period in noisy data -- more elongated asteroids are reconstructed more easily than  spherical ones. That is why almost all the asteroids listed in Tables~\ref{tab:models} and \ref{tab:models_interval} have large amplidudes of $\gtrsim 0.3$\,mag. The lightcurve inversion (based mostly or exclusively on sparse data) is also less efficient for asteroids with poles close to the ecliptic plane because, during some apparitions, we observe them almost pole-on, thus with very small amplitudes. This bias in the method was estimated to be of the order of several tens percent \citep{Han.ea:11}. A much higher discrepancy (factor 3--4) in the successfully recovered pole directions between poles close-to and perpendicular-to the ecliptic was found by \cite{San.ea:15}. But even such a large selection effect cannot fully explain the significant ``gap'' for obliquities between 60--120$^\circ$. To clearly show how the unbiased distribution of pole obliquities looks like, we would have to carry out an extensive simulation on a synthetic population with realistic systematic and random errors to see the bias that is induced by the method, shape, and geometry. This sort of simulation would be more computationally demanding than processing real data from the Lowell database. Therefore, we postpone this investigation for a future paper.

  For near-Earth asteroids, the excess of retrograde rotators can be explained by the Yarkovsky-induced delivery mechanism from the main belt through resonances \citep{LaS.ea:04}, although the number of NEA models in our sample is too small for any reliable statistics.
  
\begin{figure}
   \begin{center}
    \includegraphics[width=\columnwidth]{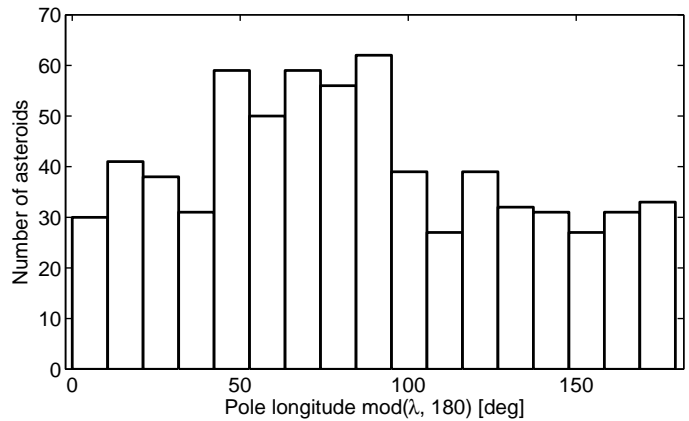}
    \caption{Histograms of the distribution of pole longitude $\lambda$ for 685 main-belt asteroids.}
    \label{fig:histogram_lambda}
   \end{center}
  \end{figure}

  We also see a clear deviation from a uniform distribution of pole longitudes in Fig.~\ref{fig:histogram_lambda}. Because of ambiguity in pole direction (often there are two solutions with similar latitudes and the difference in longitudes of about $180^\circ$), we plotted the distribution modulo $180^\circ$. The histogram shows an excess of longitudes around 50--100$^\circ$. This was already announced by \cite{Bow.ea:14}, who processed the Lowell data set using a different approach, estimated spin-axis longitudes for more than 350,000 asteroids, and revealed an excess of longitudes at 30--110$^\circ$ and a paucity at 120--180$^\circ$. The explanation of this phenomenon remains unclear.

  \section{Conclusions}
  \label{sec:conclusion}
  
  The new models presented in this paper significantly enlarge the sample of asteroids for which their spin axis direction and approximate shape are known. Because these models are based on a limited number of data points, the shapes have to be interpreted as only approximations of the real shapes of asteroids. Also the pole directions need  to be refined with more data if one is interested in a particular asteroid. However, as an ensemble, the models can be used in future statistical studies of asteroid spins, for example. 

  We believe that this is only the beginning of a mass production of shape and spin models from sparse photometry. Although the number of models derivable from the Lowell Observatory photometric database is small compared to the total number of asteroids, the potential of Lowell photometry consists in its combination with other data. Even  a priori information about the rotation period shrinks the parameter space that has to be scanned, and a local minimum in a large parameters space becomes a global minimum on a restricted interval. Of course, the reliability of this type of model depends critically on the reliability of the period. Lowell photometry can be combined with dense lightcurves that constrain the rotation period. This way, models for about 250 asteroids were derived recently by \cite{Han.ea:15b}, some of which confirm the models presented in this paper. The database of asteroid rotation periods has been increased dramatically by \cite{Was.ea:15} -- their data can also be  combined with Lowell photometry, and we expect that other hundreds of models will be reconstructed from this data set. Another promising approach is the combination of sparse photometry with data from the Wide-field Infrared Survey Explorer (WISE) mission \citep{Wri.ea:10}. Although WISE data were observed in mid-infrared wavelengths, \cite{Dur.ea:15c} showed that thermally emitted flux can be treated as reflected light to derive the correct rotation period and the shape and spin model. This opens up a new possibility, because both Lowell and WISE data are available for tens of thousands of asteroids.
  
  In general, the combination of more data sources is always better than using them separately. By using Lowell photometry with dense lightcurves, WISE data, photometry from Gaia, etc., the number of available models will increase and the statistical studies of spin and shape distribution will become more robust, being based on larger sets of models. Nevertheless, any inference based on the models derived from lightcurves (and sparse lightcurves in particular) has to take into account  that the sample of models is biased against more spherical shapes with low lightcurve amplitudes and poles near the plane of ecliptic.
  
  \begin{longtab}
    \begin{longtable}{r l r r r r @{} d @{} d l l l r c}
      \caption{\label{tab:models} List of new asteroid models derived from the full period interval 2--100 hours. For each asteroid, there is one or two pole directions in the ecliptic coordinates $(\lambda, \beta)$, the sidereal rotation period $P$, rotation period from LCDB $P_\mathrm{LCDB}$ and its quality code $U$ (if available), the minimum and maximum lightcurve amplitude $A_\text{min}$, $A_\text{max}$, respectively, the number of data points $N$, and the method which was used to derive the unique rotation period: C -- convex inversion, E -- ellipsoids, CE -- both methods gave the same unique period. The accuracy of the sidereal rotation period is of the order of the last decimal place given. Asteroids marked with $\ast$ were independently confirmed by \cite{Han.ea:15b}.}\\
      \hline\hline
      \multicolumn{2}{c}{Asteroid}              & \multicolumn{1}{c}{$\lambda_1$}       & \multicolumn{1}{c}{$\beta_1$}   & \multicolumn{1}{c}{$\lambda_2$}       & \multicolumn{1}{c}{$\beta_2$}   & \multicolumn{1}{c}{$P$}       & \multicolumn{1}{c}{$P_\mathrm{LCDB}$} & \multicolumn{1}{c}{$A_\text{min}$}      & \multicolumn{1}{c}{$A_\text{max}$}    & $U$     & \multicolumn{1}{c}{$N$}       & method        \\
      number    & name/designation              & \multicolumn{1}{c}{[deg]}             & \multicolumn{1}{c}{[deg]}       & \multicolumn{1}{c}{[deg]}             & \multicolumn{1}{c}{[deg]}       & \multicolumn{1}{c}{[h]}       & \multicolumn{1}{c}{[h]}                 & \multicolumn{1}{c}{[mag]}             & \multicolumn{1}{c}{[mag]}             &         &                               &               \\
      \hline
      \endfirsthead
      \caption{continued.}\\
      \hline\hline
      \multicolumn{2}{c}{Asteroid}              & \multicolumn{1}{c}{$\lambda_1$}       & \multicolumn{1}{c}{$\beta_1$}   & \multicolumn{1}{c}{$\lambda_2$}       & \multicolumn{1}{c}{$\beta_2$}   & \multicolumn{1}{c}{$P$}       & \multicolumn{1}{c}{$P_\mathrm{LCDB}$} & \multicolumn{1}{c}{$A_\text{min}$}      & \multicolumn{1}{c}{$A_\text{max}$}    & $U$     & \multicolumn{1}{c}{$N$}       & method        \\
      number    & name/designation              & \multicolumn{1}{c}{[deg]}             & \multicolumn{1}{c}{[deg]}       & \multicolumn{1}{c}{[deg]}             & \multicolumn{1}{c}{[deg]}       & \multicolumn{1}{c}{[h]}       & \multicolumn{1}{c}{[h]}                 & \multicolumn{1}{c}{[mag]}             & \multicolumn{1}{c}{[mag]}             &       &                               &                 \\
      \hline
      \endhead
      \hline
      \endfoot
  136 & Austria$^\ast$            &   118 &    57 &   333 &    75 & 11.49665 &     11.4969 &      & 0.37 &     3 &   401 &  CE  \\
  163 & Erigone                   &   276 & $-69$ &       &       &  16.1402 &      16.136 & 0.32 & 0.37 &     3 &   483 &   E  \\
  186 & Celuta                    &    88 & $-54$ &   235 & $-77$ &  19.8435 &      19.842 &  0.4 & 0.55 &     3 &   406 &   C  \\
  254 & Augusta$^\ast$            &    56 & $-69$ &   218 & $-75$ &  5.89503 &      5.8961 & 0.56 & 0.75 &  3$-$ &   371 &  CE  \\
  263 & Dresda$^\ast$             &   101 &    55 &   280 &    58 &  16.8138 &      16.809 & 0.32 & 0.55 &     3 &   605 &   E  \\
  274 & Philagoria$^\ast$         &   138 & $-52$ &   303 & $-58$ & 17.94072 &       17.96 & 0.43 & 0.51 &     3 &   460 &   E  \\
  296 & Phaetusa$^\ast$           &   145 &    53 &   326 &    60 & 4.538091 &      4.5385 & 0.38 & 0.50 &     3 &   340 &   C  \\
  381 & Myrrha$^\ast$             &     3 &    48 &   160 &    77 &  6.57198 &       6.572 & 0.35 & 0.36 &     3 &   496 &   E  \\
  407 & Arachne$^\ast$            &    64 & $-49$ &   268 & $-63$ &  22.6264 &       22.62 & 0.31 & 0.45 &     2 &   433 &   C  \\
  427 & Galene                    &    72 & $-57$ &   272 & $-75$ & 3.706036 &       3.705 & 0.55 & 0.68 &     3 &   394 &  CE  \\
  474 & Prudentia$^\ast$          &   150 & $-58$ &   297 & $-46$ &  8.57228 &       8.572 & 0.53 & 0.90 &     3 &   374 &   E  \\
  482 & Petrina$^\ast$            &    94 &     2 &   274 &    37 & 11.79212 &     11.7922 & 0.07 & 0.56 &     3 &   337 &   E  \\
  518 & Halawe                    &   120 & $-66$ &   292 & $-58$ & 14.31765 &      14.310 & 0.50 & 0.55 &     3 &   439 &   E  \\
  520 & Franziska$^\ast$          &   122 & $-50$ &   301 & $-59$ &  16.5044 &      16.507 & 0.35 & 0.51 &     3 &   384 &  CE  \\
  523 & Ada                       &   152 & $-70$ &   357 & $-70$ & 10.03242 &       10.03 & 0.52 & 0.70 &     3 &   343 &  CE  \\
  616 & Elly                      &    60 &    62 &   250 &    44 &  5.29770 &       5.297 & 0.34 & 0.44 &     3 &   368 &  CE  \\
  620 & Drakonia                  &   138 &    56 &   316 &    47 &  5.48711 &       5.487 & 0.52 & 0.65 &     3 &   345 &   E  \\
  632 & Pyrrha$^\ast$             &    72 & $-64$ &   249 & $-72$ & 4.116854 &      4.1167 &      & 0.40 &     3 &   487 &  CE  \\
  650 & Amalasuntha               &    46 &    51 &       &       & 16.57586 &      16.582 & 0.45 & 0.49 &     3 &   435 &   E  \\
  686 & Gersuind$^\ast$           &   127 &    56 &       &       &  6.31240 &      6.3127 & 0.30 & 0.37 &     3 &   400 &   E  \\
  689 & Zita                      &     8 & $-72$ &   256 & $-61$ &  6.42391 &       6.425 & 0.30 & 0.62 &     3 &   369 &   E  \\
  698 & Ernestina$^\ast$          &   193 & $-68$ &       &       &  5.03661 &      5.0363 & 0.30 & 0.69 &     3 &   459 &   C  \\
  718 & Erida                     &    78 & $-56$ &   257 & $-52$ &  17.4462 &      17.447 & 0.31 & 0.37 &     3 &   430 &   E  \\
  749 & Malzovia$^\ast$           &    53 &    37 &   242 &    46 &  5.92749 &      5.9279 &      & 0.30 &  3$-$ &   423 &   E  \\
  784 & Pickeringia$^\ast$        &    99 &    67 &   283 &    30 & 13.16995 &       13.17 & 0.20 & 0.40 &     2 &   437 &   E  \\
  789 & Lena                      &   192 &    39 &       &       &  5.84239 &       5.848 & 0.40 & 0.50 &     3 &   328 &   E  \\
  829 & Academia                  &    71 & $-41$ &   245 & $-67$ &  7.89321 &       7.891 & 0.36 & 0.44 &  3$-$ &   436 &   E  \\
  877 & Walkure$^\ast$            &    68 &    58 &   253 &    61 &  17.4217 &      17.424 & 0.33 & 0.44 &  3$-$ &   596 &   E  \\
  881 & Athene$^\ast$             &   123 & $-58$ &   337 & $-47$ & 13.89449 &      13.895 & 0.39 & 0.53 &  3$-$ &   376 &  CE  \\
  955 & Alstede                   &    54 &    38 &   240 &    13 &  5.18735 &        5.19 & 0.26 & 0.27 &     3 &   401 &   E  \\
  996 & Hilaritas                 &   100 & $-56$ &   281 & $-57$ & 10.05154 &       10.05 & 0.63 & 0.70 &     3 &   442 &  CE  \\
  998 & Bodea                     &     7 & $-59$ &       &       &  8.57412 &       8.574 &      & 0.68 &     3 &   262 &   E  \\
 1017 & Jacqueline                &     7 &    55 &   170 &    65 &  7.87149 &        7.87 &  0.6 & 0.72 &     3 &   491 &  CE  \\
 1035 & Amata                     &    31 &    69 &   247 &    29 &  9.08215 &       9.081 & 0.41 & 0.44 &     3 &   305 &   E  \\
 1050 & Meta                      &    60 & $-42$ &   198 & $-79$ &  6.14188 &       6.142 &      & 0.46 &     3 &   325 &   E  \\
 1061 & Paeonia                   &   155 & $-50$ &       &       &  7.99710 &          6. &      &  0.5 &  2$-$ &   314 &   E  \\
 1075 & Helina                    &   123 & $-33$ &   284 & $-34$ &  44.6768 &        44.9 &      & 0.64 &  3$-$ &   421 &   C  \\
 1081 & Reseda                    &    92 & $-69$ &   256 & $-76$ &  7.30136 &      7.3002 &      & 0.34 &     3 &   410 &   E  \\
 1082 & Pirola                    &   123 & $-42$ &   300 & $-38$ &  15.8540 &     15.8525 & 0.53 & 0.60 &     3 &   528 &   E  \\
 1098 & Hakone                    &    40 &    43 &       &       &  7.14117 &       7.142 & 0.35 & 0.40 &     3 &   382 &   E  \\
 1119 & Euboea$^\ast$             &    71 &    61 &   280 &    54 & 11.39823 &       11.41 & 0.46 & 0.50 &     3 &   461 &   E  \\
 1121 & Natascha                  &    16 &    59 &   209 &    50 & 13.19717 &      13.197 &      & 0.51 &     3 &   415 &   E  \\
 1127 & Mimi                      &   224 & $-57$ &       &       & 12.74557 &      12.749 & 0.72 & 0.95 &     3 &   357 &  CE  \\
 1135 & Colchis$^\ast$            &     7 & $-54$ &   168 & $-56$ &  23.4827 &       23.47 &      & 0.45 &     2 &   409 &   C  \\
 1147 & Stavropolis               &    78 & $-50$ &   267 & $-51$ &  5.66079 &        5.66 &      & 0.42 &     3 &   372 &   E  \\
 1187 & Afra                      &    40 &    34 &   226 &    13 & 14.06993 &     14.0701 & 0.38 & 0.40 &     3 &   374 &   E  \\
 1204 & Renzia$^\ast$             &   130 & $-44$ &   312 & $-51$ &  7.88697 &       7.885 &      & 0.42 &     3 &   528 &   E  \\
 1206 & Numerowia                 &    64 & $-50$ &   271 & $-69$ &  4.77529 &      4.7743 &      & 0.63 &     3 &   322 &  CE  \\
 1219 & Britta                    &    72 & $-66$ &   241 & $-66$ &  5.57556 &       5.575 & 0.48 & 0.75 &     3 &   387 &   E  \\
 1230 & Riceia                    &    37 & $-63$ &       &       &  6.67317 &             &      &      &       &   293 &  CE  \\
 1231 & Auricula                  &    57 & $-57$ &   225 & $-85$ & 3.981580 &      3.9816 &      & 0.75 &     3 &   292 &   E  \\
 1245 & Calvinia                  &    52 & $-51$ &   235 & $-43$ &  4.85148 &        4.84 & 0.37 &  0.7 &     3 &   410 &   E  \\
 1248 & Jugurtha                  &   254 & $-89$ &       &       & 12.19047 &      12.910 & 0.70 &  1.4 &     3 &   381 &   C  \\
 1251 & Hedera                    &   124 & $-70$ &   266 & $-62$ &  19.9020 &     19.9000 & 0.41 & 0.61 &  3$-$ &   415 &   E  \\
 1275 & Cimbria                   &    85 & $-61$ &   271 & $-31$ &  5.65454 &        5.65 & 0.40 & 0.57 &     3 &   352 &   E  \\
 1281 & Jeanne                    &   153 &    19 &   338 &    32 & 15.30379 &        15.2 &      & 0.45 &     2 &   470 &   E  \\
 1299 & Mertona                   &    73 &    35 &   253 &    56 &  4.97691 &       4.977 & 0.46 & 0.59 &     3 &   369 &   E  \\
 1312 & Vassar$^\ast$             &   104 & $-50$ &   250 & $-29$ &  7.93189 &       7.932 &      & 0.35 &     3 &   317 &   E  \\
 1320 & Impala                    &   151 & $-57$ &   254 & $-70$ &  6.17081 &       6.167 & 0.40 & 0.52 &    2+ &   353 &   E  \\
 1334 & Lundmarka                 &    79 & $-75$ &       &       &  6.25033 &       6.250 &      & 0.70 &  3$-$ &   496 &  CE  \\
 1339 & Desagneauxa               &    63 &    53 &   225 &    42 &  9.37514 &       9.380 & 0.45 & 0.48 &     3 &   465 &   E  \\
 1349 & Bechuana                  &   153 &    32 &   314 &    46 &  15.6873 &      15.692 &      & 0.30 &  3$-$ &   412 &   E  \\
 1350 & Rosselia                  &    67 & $-64$ &   246 & $-58$ &  8.14008 &       8.140 &  0.3 & 0.54 &     3 &   425 &   E  \\
 1391 & Carelia                   &    21 & $-79$ &   208 & $-43$ &  5.87822 &             &      &      &       &   295 &   E  \\
 1400 & Tirela                    &    58 & $-80$ &   297 & $-41$ & 13.35384 &      13.356 &      & 0.55 &     2 &   281 &   E  \\
 1459 & Magnya$^\ast$             &    73 & $-54$ &   198 & $-55$ & 4.679102 &       4.678 & 0.57 & 0.85 &     3 &   363 &   E  \\
 1484 & Postrema                  &    19 &    44 &   250 &    64 & 12.18978 &     12.1923 & 0.22 & 0.23 &  3$-$ &   312 &   E  \\
 1493 & Sigrid                    &   183 &    69 &   350 &    69 &  43.1795 &      43.296 & 0.38 &  0.6 &     2 &   452 &   C  \\
 1494 & Savo                      &    50 & $-65$ &   233 & $-68$ &  5.35059 &     5.35011 & 0.45 & 0.52 &     3 &   486 &   C  \\
 1500 & Jyvaskyla                 &   123 & $-75$ &   268 & $-79$ &  8.82750 &             &      &      &       &   248 &   C  \\
 1545 & Thernoe                   &   164 & $-75$ &   352 & $-80$ & 17.20321 &       17.20 &      & 0.76 &     3 &   281 &   E  \\
 1547 & Nele                      &   159 &    28 &   318 &    50 &  7.09742 &       7.100 & 0.18 & 0.45 &  3$-$ &   343 &   E  \\
 1548 & Palomaa                   &    72 & $-61$ &   232 & $-32$ &  7.49966 &      7.4961 &      & 0.50 &     3 &   353 &   E  \\
 1551 & Argelander                &     3 & $-81$ &   183 & $-72$ & 4.058350 &             &      &      &       &   453 &  CE  \\
 1557 & Roehla                    &   124 & $-38$ &   329 & $-57$ &  5.67899 &             &      &      &       &   334 &  CE  \\
 1561 & Fricke                    &   320 &    71 &       &       & 15.15330 &             &      &      &       &   395 &   E  \\
 1597 & Laugier                   &   345 & $-78$ &       &       &  8.02272 &             &      &      &       &   321 &   E  \\
 1619 & Ueta                      &    99 &    49 &   295 &    37 & 2.718238 &       2.720 & 0.32 & 0.44 &     3 &   350 &   E  \\
 1623 & Vivian                    &    52 & $-66$ &   229 & $-56$ &  20.5235 &     20.5209 &      & 0.85 &  3$-$ &   316 &   C  \\
 1643 & Brown                     &   140 &    64 &   353 &    84 &  5.93124 &       5.932 &      & 0.48 &     3 &   497 &  CE  \\
 1648 & Shajna$^\ast$             &    94 &    47 &   278 &    50 &  6.41368 &      6.4140 &      & 0.65 &     3 &   391 &   E  \\
 1672 & Gezelle$^\ast$            &    44 &    73 &   234 &    82 &  40.6821 &       40.72 &  0.2 & 0.56 &     3 &   366 &   C  \\
 1676 & Kariba$^\ast$             &    71 &    74 &   279 &    56 & 3.167336 &      3.1673 & 0.51 & 0.65 &     3 &   342 &  CE  \\
 1687 & Glarona                   &   132 &    76 &   274 &    70 &  6.49595 &         6.3 &      & 0.75 &     3 &   375 &  CE  \\
 1730 & Marceline$^\ast$          &    95 &    56 &   303 &    81 & 3.836550 &       3.837 & 0.94 & 1.00 &     3 &   268 &   E  \\
 1733 & Silke                     &   141 & $-63$ &   302 & $-58$ &  7.89457 &             &      &      &       &   277 &   E  \\
 1738 & Oosterhoff                &   121 & $-62$ &   301 & $-80$ & 4.448955 &      4.4486 & 0.48 & 0.54 &     3 &   371 &   C  \\
 1743 & Schmidt                   &    69 & $-62$ &   261 & $-53$ &  17.4599 &       17.45 &      & 0.36 &     3 &   381 &   E  \\
 1753 & Mieke                     &   121 &    67 &   321 &    35 & 10.19942 &         8.8 &      &  0.2 &     2 &   413 &   E  \\
 1758 & Naantali                  &   150 & $-60$ &       &       &  5.47369 &      5.4699 &      & 0.44 &     3 &   446 &   E  \\
 1768 & Appenzella                &    39 &    45 &   227 &    40 &  5.18335 &      5.1839 &      & 0.53 &     3 &   395 &  CE  \\
 1774 & Kulikov                   &    54 & $-50$ &   237 & $-46$ & 3.830791 &             &      &      &       &   518 &   E  \\
 1792 & Reni                      &   122 & $-41$ &   260 & $-50$ & 15.94121 &       15.95 &      & 0.54 &  3$-$ &   319 &   E  \\
 1793 & Zoya$^\ast$               &    52 &    59 &   227 &    59 &  5.75187 &       5.753 &      & 0.40 &    2+ &   398 &  CE  \\
 1801 & Titicaca                  &    48 &    51 &   260 &    57 & 3.211233 &      3.2106 &      & 0.50 &     3 &   379 &  CE  \\
 1814 & Bach                      &   126 &    66 &   317 &    64 &  7.23954 &             &      &      &       &   221 &  CE  \\
 1819 & Laputa                    &   149 & $-48$ &       &       &  9.79965 &      9.8004 &      & 0.51 &     3 &   394 &  CE  \\
 1825 & Klare$^\ast$              &   115 & $-61$ &   315 & $-69$ & 4.742885 &       4.744 & 0.70 & 0.90 &     3 &   336 &  CE  \\
 1838 & Ursa$^\ast$               &    51 &    66 &   286 &    32 & 16.16358 &      16.141 &      & 0.80 &     3 &   346 &   C  \\
 1840 & Hus                       &   298 & $-77$ &       &       & 4.749057 &       4.780 &      & 0.85 &  2$-$ &   549 &   E  \\
 1841 & Masaryk                   &   122 &    62 &   305 &    59 &  7.54301 &        7.53 &      & 0.52 &    2+ &   406 &   E  \\
 1855 & Korolev                   &    90 &    52 &   262 &    64 & 4.656199 &      4.6568 & 0.75 & 0.76 &     3 &   370 &   C  \\
 1900 & Katyusha                  &    94 & $-46$ &   291 & $-48$ &  9.50358 &      9.4999 & 0.56 & 0.74 &     3 &   283 &   E  \\
 1942 & Jablunka                  &   156 & $-73$ &       &       &  8.91158 &             &      &      &       &   225 &   E  \\
 1945 & Wesselink                 &   190 & $-78$ &   336 & $-60$ & 3.547454 &             &      &      &       &   445 &   E  \\
 1949 & Messina                   &   138 & $-64$ &   326 & $-51$ & 3.649308 &      3.6491 &      & 0.37 &     3 &   323 &  CE  \\
 1978 & Patrice                   &    21 &    13 &   203 &    18 & 5.881213 &             &      &      &       &   365 &  CE  \\
 1985 & Hopmann                   &   107 & $-81$ &       &       &  17.4787 &      17.480 & 0.36 & 0.44 &     3 &   272 &   E  \\
 1997 & Leverrier                 &    96 &    46 &   274 &    40 &  8.01532 &             &      &      &       &   301 &   E  \\
 2313 & Aruna$^\ast$              &    94 & $-80$ &   283 & $-70$ &  8.88618 &        8.90 &      & 0.79 &     3 &   715 &   E  \\
 2381 & Landi$^\ast$              &    54 & $-87$ &   237 & $-45$ & 3.986045 &       3.989 & 0.75 & 1.04 &     3 &   364 &   E  \\
 2395 & Aho                       &   160 &    47 &   340 &    46 &  7.88033 &             &      &      &       &   751 &   E  \\
 2425 & Shenzhen                  &    50 &    58 &   265 &    40 &  9.83818 &      14.715 &      & 0.80 &     3 &   551 &   E  \\
 2483 & Guinevere                 &    19 &    70 &   194 &    59 & 14.73081 &      14.733 & 1.34 & 1.38 &     3 &   317 &   E  \\
 2528 & Mohler                    &    56 & $-64$ &   246 & $-54$ &  6.49130 &             &      &      &       &   641 &   E  \\
 2581 & Radegast                  &    57 &    56 &   230 &    53 &  8.75121 &             &      &      &       &   573 &  CE  \\
 2630 & Hermod                    &    74 &    50 &   256 &    45 &  19.4283 &             &      &      &       &   578 &   E  \\
 2659 & Millis$^\ast$             &   117 & $-55$ &   294 & $-52$ &  6.12464 &       6.132 & 0.53 & 0.84 &     3 &   566 &   C  \\
 2836 & Sobolev                   &    82 & $-51$ &   270 & $-79$ & 4.754883 &             &      &      &       &   560 &   E  \\
 3086 & Kalbaugh                  &    63 & $-51$ &       &       &  5.17907 &       5.180 & 0.47 & 0.76 &     3 &   383 &   C  \\
 3261 & Tvardovskij               &    90 &    65 &   280 &    68 &  5.36852 &             &      &      &       &   665 &   E  \\
 3281 & Maupertuis                &    62 & $-66$ &   231 & $-74$ &  6.72984 &      6.7295 &      & 1.22 &     3 &   453 &   E  \\
 3286 & Anatoliya                 &    52 &    51 &   293 &    76 &  5.81029 &             &      &      &       &   410 &  CE  \\
 3375 & Amy                       &   168 & $-50$ &   343 & $-49$ & 3.255633 &             &      &      &       &   486 &   E  \\
 3407 & Jimmysimms                &    34 &    53 &   259 &    82 &  6.82069 &       6.819 & 0.93 & 0.95 &     3 &   457 &   E  \\
 3544 & Borodino$^\ast$           &   104 & $-57$ &   267 & $-53$ &  5.43459 &       5.442 & 0.60 & 0.65 &     3 &   515 &   E  \\
 3573 & Holmberg                  &   142 &    50 &   318 &    52 &  6.54245 &      6.5431 & 0.91 & 1.03 &     3 &   577 &   C  \\
 3735 & Trebon                    &    51 & $-54$ &   236 & $-69$ &  8.47251 &             &      &      &       &   568 &   E  \\
 3746 & Heyuan                    &   170 &    66 &   333 &    70 &  16.3010 &             &      &      &       &   538 &   E  \\
 3758 & Karttunen                 &    74 & $-72$ &   202 & $-52$ & 12.50101 &             &      &      &       &   396 &  CE  \\
 3786 & Yamada$^\ast$             &   100 &    54 &   205 &    60 & 4.032946 &       4.034 & 0.40 & 0.65 &     3 &   463 &   E  \\
 3822 & Segovia                   &    43 &    58 &   265 &    72 & 11.03204 &             &      &      &       &   585 &   E  \\
 3910 & Liszt                     &   104 & $-46$ &   290 & $-66$ & 4.736280 &        4.73 &      & 0.60 &     3 &   423 &   E  \\
 4037 & Ikeya                     &    92 &    67 &   270 &    44 & 4.057537 &             &      &      &       &   485 &   E  \\
 4877 & Humboldt                  &   218 & $-77$ &   340 & $-55$ & 3.491213 &             &      &      &       &   450 &   E  \\
 5006 & Teller                    &    84 &    66 &   301 &    57 & 10.90225 &      10.898 &      & 0.69 &     3 &   474 &   E  \\
 5195 & Kaendler                  &    67 & $-59$ &   212 & $-50$ &  5.33756 &             &      &      &       &   537 &   E  \\
 5299 & Bittesini                 &    76 &    60 &   251 &    49 & 4.679660 &             &      &      &       &   599 &  CE  \\
 5488 & Kiyosato                  &    19 &    23 &   242 &    62 &  8.76307 &             &      &      &       &   449 &   E  \\
 5489 & Oberkochen$^\ast$         &    23 & $-62$ &   194 & $-38$ &  5.62439 &       5.625 & 0.40 & 0.51 &     3 &   470 &   E  \\
 5494 & 1933 UM1                  &   137 & $-65$ &   323 & $-62$ &  5.72752 &             &      &      &       &   612 &  CE  \\
 5685 & Sanenobufukui             &   122 & $-63$ &   327 & $-58$ & 3.387871 &             &      &      &       &   518 &  CE  \\
 5723 & Hudson                    &    72 & $-73$ &   255 & $-58$ & 4.475115 &             &      &      &       &   472 &   E  \\
 5776 & 1989 UT2$^\ast$           &   105 & $-76$ &   350 & $-46$ & 4.340787 &             &      &      &       &   473 &   E  \\
 5929 & 1974 XT                   &   130 & $-71$ &   244 & $-40$ & 3.759432 &      3.7596 &      & 1.01 &     3 &   358 &   E  \\
 5993 & Tammydickinson            &    52 & $-65$ &   241 & $-78$ &  9.44711 &             &      &      &       &   546 &   E  \\
 6136 & Gryphon                   &   134 &    57 &   337 &    63 &  16.4683 &      16.476 &      & 0.61 &     3 &   602 &   E  \\
 6161 & Vojno-Yasenetsky          &    64 &    70 &   217 &    41 &  7.98095 &             &      &      &       &   393 &   E  \\
 6166 & Univsima                  &    80 & $-42$ &   242 & $-59$ & 11.37655 &         9.6 &      &  0.6 &     2 &   425 &   E  \\
 6276 & Kurohone                  &   147 &    63 &   322 &    62 &  6.34850 &             &      &      &       &   441 &   E  \\
 6410 & Fujiwara$^\ast$           &   148 & $-59$ &   296 & $-81$ &  7.00667 &      7.0073 & 0.80 & 0.85 &     3 &   552 &   E  \\
 6422 & Akagi                     &   213 & $-38$ &       &       &  7.74756 &             &      &      &       &   430 &   E  \\
 6590 & Barolo                    &   171 &    59 &   313 &    63 &  8.35928 &             &      &      &       &   460 &  CE  \\
 6671 & 1994 NC1                  &    69 &    57 &   204 &    78 &  5.22042 &             &      &      &       &   463 &   E  \\
 6719 & Gallaj                    &    60 & $-63$ &   252 & $-59$ & 4.429487 &             &      &      &       &   515 &   E  \\
 6882 & Sormano                   &    43 & $-33$ &   248 & $-58$ & 3.998344 &             &      &      &       &   396 &   E  \\
 7001 & Noether                   &    13 & $-66$ &       &       &  9.58191 &             &      &      &       &   421 &   E  \\
 7072 & Beijingdaxue              &    72 &    56 &   262 &    64 & 5.304419 &             &      &      &       &   549 &   E  \\
 7106 & Kondakov                  &    63 &    58 &   268 &    81 &  7.59690 &             &      &      &       &   515 &   E  \\
 7289 & Kamegamori                &   108 & $-39$ &   268 & $-54$ & 3.831182 &             &      &      &       &   538 &   C  \\
 7318 & Dyukov                    &    65 & $-64$ &   239 & $-44$ & 4.856335 &             &      &      &       &   497 &   E  \\
 7835 & 1993 MC                   &    72 & $-64$ &   288 & $-55$ &  7.43019 &             &      &      &       &   473 &   E  \\
 7896 & Svejk                     &    86 &    36 &   266 &    35 & 16.20582 &             &      &      &       &   500 &   E  \\
 7964 & 1995 DD2                  &    74 & $-73$ &   264 & $-49$ & 10.22553 &             &      &      &       &   532 &   E  \\
 8573 & Ivanka                    &   100 & $-70$ &   344 & $-78$ &  8.03312 &             &      &      &       &   454 &   E  \\
 9440 & 1997 FZ1                  &    97 & $-64$ &   278 & $-65$ & 5.196349 &             &      &      &       &   485 &   E  \\
 9971 & Ishihara                  &    42 &    76 &   223 &    60 &  6.71574 &             &      &      &       &   559 &   E  \\
10281 & 1981 EE45                 &   183 & $-88$ &       &       &  7.57192 &             &      &      &       &   462 &   E  \\
10472 & 1981 EO20                 &    73 &    37 &   249 &    43 &  5.98599 &             &      &      &       &   344 &   C  \\
10627 & Ookuninushi               &     3 &    49 &   204 &    89 & 4.334986 &             &      &      &       &   394 &   E  \\
11052 & 1990 WM                   &   194 & $-49$ &   351 & $-78$ &  5.06904 &             &      &      &       &   471 &   E  \\
11148 & Einhardress               &   121 & $-69$ &   296 & $-64$ &  7.77240 &             &      &      &       &   378 &   E  \\
11700 & 1998 FT115                &    82 & $-64$ &   280 & $-72$ &  5.52303 &             &      &      &       &   596 &   E  \\
11958 & Galiani                   &    12 &    47 &       &       &  8.24720 &       9.801 &      & 0.96 &    2+ &   541 &   E  \\
11995 & 1995 YB1                  &    70 & $-80$ &   238 & $-68$ & 12.66498 &             &      &      &       &   398 &   E  \\
12384 & Luigimartella             &     4 & $-50$ &   160 & $-60$ &  6.44220 &             &      &      &       &   517 &   E  \\
12551 & 1998 QQ39                 &   123 &    50 &   299 &    44 & 11.19841 &             &      &      &       &   571 &   E  \\
12753 & Povenmire                 &   138 & $-28$ &   295 & $-48$ &  17.5740 &      12.854 &      & 0.45 &     2 &   424 &   E  \\
12774 & Pfund                     &   121 &    53 &   308 &    62 &  7.34295 &             &      &      &       &   174 &   E  \\
12979 & 1978 SB8                  &    45 & $-68$ &   203 & $-62$ &  5.78331 &             &      &      &       &   349 &   C  \\
13059 & Ducuroir                  &   117 &    71 &   311 &    73 & 13.84864 &             &      &      &       &   560 &   E  \\
13338 & 1998 SK119                &    86 & $-42$ &   240 & $-70$ & 4.129105 &             &      &      &       &   474 &   E  \\
13535 & 1991 RS13                 &   127 & $-64$ &   318 & $-60$ &  5.22705 &             &      &      &       &   390 &   E  \\
13952 & 1990 SN6                  &    13 &    58 &   193 &    67 &  7.35225 &             &      &      &       &   526 &   E  \\
14031 & 1994 WF2                  &    41 & $-62$ &   180 & $-60$ & 2.900652 &             &      &      &       &   509 &   E  \\
14044 & 1995 VS1                  &    95 & $-47$ &   253 & $-57$ &  9.25874 &             &      &      &       &   499 &   E  \\
14203 & Hocking                   &   106 &    81 &   240 &    52 &  9.08566 &             &      &      &       &   393 &   E  \\
14691 & 2000 AK119                &    38 & $-54$ &   248 & $-68$ & 3.652412 &       3.652 & 0.70 & 0.78 &     3 &   441 &   E  \\
15677 & 1980 TZ5                  &    75 & $-64$ &   226 & $-75$ &  5.61479 &             &      &      &       &   491 &   E  \\
16216 & 2000 DR4                  &   130 & $-66$ &       &       & 10.36816 &             &      &      &       &   480 &   E  \\
16786 & 1997 AT1                  &   178 &    56 &   307 &    60 & 4.020902 &             &      &      &       &   296 &   E  \\
16955 & 1998 KU48                 &   122 & $-48$ &   264 & $-51$ &  5.26062 &             &      &      &       &   387 &   E  \\
17111 & 1999 JH52                 &   157 &    46 &   328 &    62 &  8.89847 &             &      &      &       &   629 &   E  \\
19608 & 1999 NC57                 &    69 & $-68$ &   306 & $-68$ &  9.18995 &             &      &      &       &   513 &   E  \\
20329 & Manfro                    &    92 & $-55$ &   258 & $-63$ &  8.80724 &             &      &      &       &   395 &   E  \\
20570 & Molchan                   &    40 & $-52$ &   244 & $-59$ & 4.132007 &             &      &      &       &   485 &   E  \\
20725 & 1999 XP120                &   170 &    67 &   334 &    45 & 3.601296 &             &      &      &       &   488 &   E  \\
21411 & Abifraeman                &   249 & $-63$ &       &       & 11.02020 &             &      &      &       &   409 &   E  \\
22018 & 1999 XK105                &    61 &    45 &   228 &    31 &  17.0575 &             &      &      &       &   373 &   C  \\
22298 & 1990 EJ                   &    64 & $-42$ &   244 & $-72$ & 2.985538 &             &      &      &       &   530 &   C  \\
23578 & Baedeker                  &   141 & $-53$ &       &       &  8.17154 &             &      &      &       &   382 &   C  \\
23707 & 1997 TZ7                  &   185 &    79 &   334 &    35 &  5.06632 &             &      &      &       &   381 &   E  \\
23873 & 1998 RL76                 &    56 & $-74$ &       &       & 13.86111 &             &      &      &       &   323 &   E  \\
26241 & 1998 QY40                 &    40 &    45 &       &       & 12.83303 &             &      &      &       &   380 &   C  \\
26387 & 1999 TG2                  &    94 & $-46$ &   325 & $-79$ &  5.14720 &             &      &      &       &   460 &   E  \\
26460 & 2000 AZ120                &     9 & $-87$ &   230 & $-47$ &  5.45444 &        5.48 &      & 0.90 &     2 &   326 &   E  \\
27225 & 1999 GB17                 &    72 &    54 &   266 &    31 & 3.853858 &             &      &      &       &   480 &   C  \\
28133 & 1998 SS130                &   126 & $-49$ &   310 & $-56$ &  5.46284 &             &      &      &       &   408 &  CE  \\
29308 & 1993 UF1                  &    71 & $-56$ &   246 & $-83$ &  9.79348 &       9.810 & 0.83 & 0.94 &     3 &   365 &   E  \\
29777 & 1999 CK46                 &    32 &    56 &   224 &    40 & 3.627265 &             &      &      &       &   460 &   E  \\
31060 & 1996 TB6$^\ast$           &   103 & $-52$ &   253 & $-76$ & 5.104323 &       5.103 &      & 0.80 &     3 &   394 &   E  \\
32575 & 2001 QY78                 &    60 & $-70$ &   223 & $-64$ & 4.535495 &      4.5344 &      & 0.86 &     3 &   324 &   E  \\
32799 & 1990 QN1                  &    71 &    75 &   258 &    72 &  9.49636 &             &      &      &       &   337 &   E  \\
33776 & 1999 RB158                &    46 &    64 &   247 &    42 &  16.9593 &             &      &      &       &   426 &   E  \\
33854 & 2000 HH53                 &   114 & $-56$ &   236 & $-78$ & 4.423484 &             &      &      &       &   348 &   E  \\
33974 & 2000 ND17                 &   132 & $-55$ &   347 & $-65$ &  4.67868 &             &      &      &       &   340 &   E  \\
34318 & 2000 QV192                &    81 & $-70$ &   237 & $-36$ &  6.35105 &             &      &      &       &   198 &   E  \\
35218 & 1994 WU2                  &   138 & $-56$ &   304 & $-68$ & 11.56991 &             &      &      &       &   353 &   E  \\
35928 & 1999 JV107                &   137 & $-52$ &   302 & $-75$ &  7.26361 &             &      &      &       &   279 &   E  \\
35965 & 1999 LH13                 &    63 & $-70$ &   245 & $-55$ &  6.76072 &             &      &      &       &   387 &   E  \\
36303 & 2000 JM54                 &    55 & $-48$ &   225 & $-46$ &  4.93652 &             &      &      &       &   368 &   E  \\
36487 & 2000 QJ42                 &   127 & $-66$ &   332 & $-50$ &  8.14982 &             &      &      &       &   333 &   E  \\
36944 & 2000 SD249                &    96 & $-69$ &   295 & $-76$ &  6.09677 &             &      &      &       &   188 &   C  \\
40104 & 1998 QE4                  &   144 & $-87$ &       &       & 4.475241 &             &      &      &       &   443 &   E  \\
40478 & 1999 RT54                 &   184 & $-66$ &   358 & $-65$ & 3.886734 &             &      &      &       &   286 &   C  \\
40806 & 1999 TX41                 &    82 & $-69$ &   269 & $-64$ &  6.41500 &             &      &      &       &   198 &   C  \\
43163 & 1999 XB127                &   154 & $-59$ &       &       &  5.20660 &             &      &      &       &   298 &   C  \\
43574 & 2001 FU192                &   104 & $-36$ &   279 & $-49$ & 4.324751 &             &      &      &       &   239 &   E  \\
45864 & 2000 UO97                 &    44 & $-66$ &   179 & $-84$ &  5.13544 &             &      &      &       &   436 &   E  \\
46376 & 2001 XD3                  &   121 &    55 &       &       &  5.73569 &             &      &      &       &   278 &   E  \\
47508 & 2000 AQ58                 &    12 & $-49$ &   196 & $-26$ &  5.05521 &             &      &      &       &   349 &   E  \\
48268 & 2002 AK1                  &    72 & $-69$ &   234 & $-35$ & 4.250546 &             &      &      &       &   406 &   E  \\
48842 & 1998 BA44                 &    86 &    83 &   254 &    43 &  6.17996 &             &      &      &       &   383 &   E  \\
52723 & 1998 GP2                  &    86 & $-39$ &   224 & $-56$ & 11.90781 &             &      &      &       &   329 &   E  \\
54850 & 2001 OZ11                 &    20 & $-68$ &   189 & $-48$ & 4.222157 &             &      &      &       &   283 &   E  \\
55200 & 2001 RO19                 &    14 & $-70$ &   186 & $-58$ &  18.3910 &             &      &      &       &   233 &   E  \\
64480 & 2001 VG45                 &   128 & $-60$ &   265 & $-72$ &  6.49717 &             &      &      &       &   233 &   E  \\
66076 & 1998 RD53                 &   100 & $-54$ &   233 & $-66$ &  8.93855 &             &      &      &       &   485 &   E  \\
69117 & 2003 EX2                  &    13 &    70 &       &       &  4.74608 &             &      &      &       &   215 &   E  \\
71011 & 1999 XE45                 &    91 & $-24$ &   264 & $-31$ &  4.69449 &             &      &      &       &   115 &   C  \\
74155 & 1998 QK93                 &   264 &    50 &       &       &  15.9680 &             &      &      &       &   393 &   E  \\
75167 & 1999 VF128                &   190 & $-56$ &   350 & $-55$ & 10.27743 &             &      &      &       &   280 &   C  \\
75495 & 1999 XM181                &    64 & $-32$ &   252 & $-64$ & 14.69041 &             &      &      &       &   200 &   E  \\
76214 & 2000 EV64                 &   168 &    28 &       &       &  9.27775 &             &      &      &       &   132 &   C  \\
77677 & 2001 MA25                 &   139 & $-69$ &   326 & $-55$ & 5.304712 &             &      &      &       &   253 &   E  \\
78420 & 2002 QU40                 &   123 & $-46$ &   327 & $-45$ &  4.91365 &        4.90 &      &  1.1 &     2 &   193 &   E  \\
79436 & 1997 TD6                  &    70 & $-53$ &   201 & $-68$ & 3.635352 &             &      &      &       &   207 &   E  \\
80112 & 1999 RN61                 &    70 & $-58$ &   247 & $-39$ & 4.923622 &             &      &      &       &   157 &   E  \\
81740 & 2000 JA46                 &   100 & $-72$ &   310 & $-47$ &  9.34151 &             &      &      &       &   272 &   E  \\
81911 & 2000 NV9                  &   136 &    23 &   334 &    49 &  6.95480 &             &      &      &       &   288 &   E  \\
82642 & 2001 PX5                  &    53 &    37 &   241 &    67 & 11.61144 &             &      &      &       &   178 &   E  \\
85489 & 1997 SV2                  &   317 & $-56$ &       &       &  5.79502 &             &      &      &       &    81 &   E  \\
85532 & 1997 WD21                 &   117 & $-75$ &       &       &  4.76820 &             &      &      &       &   230 &   E  \\
89764 & 2002 AW61                 &     8 & $-45$ &   152 & $-60$ &  5.93513 &             &      &      &       &   278 &   E  \\
91063 & 1998 FX62                 &    53 & $-44$ &   191 & $-60$ & 11.66510 &             &      &      &       &   255 &   C  \\
94808 & 2001 XM167                &     9 & $-59$ &   197 & $-63$ & 10.66092 &             &      &      &       &   104 &   E  \\
96461 & 1998 HS36                 &   166 & $-60$ &   344 & $-72$ & 12.67641 &             &      &      &       &   115 &   E  \\
97346 & 2000 AF10                 &    11 &    42 &   222 &    50 &  7.65364 &             &      &      &       &   394 &   E  \\
99667 & 2002 JO1                  &    86 & $-57$ &   262 & $-42$ &  5.07877 &             &      &      &       &   199 &   C  \\
    \end{longtable}
  \end{longtab}

  \begin{longtab}
    \begin{longtable}{r l r r r r @{} d @{} d l l l r c}
      \caption{\label{tab:models_interval} List of new asteroid models derived from a restricted period interval centered at $P_\mathrm{LCDB}$. The meaning of columns is the same as in Table~\ref{tab:models}. Asteroids marked with $\ast$ were independently confirmed by \cite{Han.ea:15b}.}\\
      \hline\hline
      \multicolumn{2}{c}{Asteroid}              & \multicolumn{1}{c}{$\lambda_1$}       & \multicolumn{1}{c}{$\beta_1$}   & \multicolumn{1}{c}{$\lambda_2$}       & \multicolumn{1}{c}{$\beta_2$}   & \multicolumn{1}{c}{$P$}       & \multicolumn{1}{c}{$P_\mathrm{LCDB}$} & \multicolumn{1}{c}{$A_\text{min}$}      & \multicolumn{1}{c}{$A_\text{max}$}    & $U$     & \multicolumn{1}{c}{$N$}       & method        \\
      number    & name/designation              & \multicolumn{1}{c}{[deg]}             & \multicolumn{1}{c}{[deg]}       & \multicolumn{1}{c}{[deg]}             & \multicolumn{1}{c}{[deg]}       & \multicolumn{1}{c}{[h]}       & \multicolumn{1}{c}{[h]}                 & \multicolumn{1}{c}{[mag]}             & \multicolumn{1}{c}{[mag]}             &       &                               &                 \\
      \hline
      \endfirsthead
      \caption{continued.}\\
      \hline\hline
      \multicolumn{2}{c}{Asteroid}              & \multicolumn{1}{c}{$\lambda_1$}       & \multicolumn{1}{c}{$\beta_1$}   & \multicolumn{1}{c}{$\lambda_2$}       & \multicolumn{1}{c}{$\beta_2$}   & \multicolumn{1}{c}{$P$}       & \multicolumn{1}{c}{$P_\mathrm{LCDB}$} & \multicolumn{1}{c}{$A_\text{min}$}      & \multicolumn{1}{c}{$A_\text{max}$}    & $U$     & \multicolumn{1}{c}{$N$}       & method        \\
      number    & name/designation              & \multicolumn{1}{c}{[deg]}             & \multicolumn{1}{c}{[deg]}       & \multicolumn{1}{c}{[deg]}             & \multicolumn{1}{c}{[deg]}       & \multicolumn{1}{c}{[h]}       & \multicolumn{1}{c}{[h]}                & \multicolumn{1}{c}{[mag]}             & \multicolumn{1}{c}{[mag]}             &       &                               &                 \\
      \hline
      \endhead
      \hline
      \endfoot
   60 & Echo                      &    91 & $-25$ &   272 & $-17$ &  25.2285 &      25.208 & 0.07 & 0.22 &     3 &   446 &   E  \\
  116 & Sirona                    &    51 & $-53$ &   222 & $-53$ & 12.03251 &      12.028 & 0.42 & 0.55 &     3 &   454 &   C  \\
  138 & Tolosa                    &    49 & $-42$ &   222 & $-39$ & 10.10306 &      10.101 & 0.18 & 0.45 &     3 &   528 &   E  \\
  176 & Iduna                     &    83 &    24 &   219 &    68 & 11.28784 &     11.2877 & 0.14 & 0.43 &     3 &   491 &   E  \\
  214 & Aschera                   &   123 & $-37$ &   306 & $-42$ &  6.83369 &       6.835 & 0.20 & 0.22 &     3 &   386 &   E  \\
  239 & Adrastea                  &   226 & $-70$ &       &       &  18.4717 &     18.4707 & 0.34 & 0.51 &     3 &   416 &   E  \\
  270 & Anahita$^\ast$            &    30 & $-35$ &   205 & $-48$ & 15.05950 &       15.06 & 0.25 & 0.34 &     3 &   492 &   C  \\
  353 & Ruperto-Carola$^\ast$     &    47 & $-55$ &   220 & $-46$ & 2.738962 &     2.73898 &      & 0.32 &     3 &   368 &   E  \\
  391 & Ingeborg$^\ast$           &   305 & $-52$ &       &       &  26.4149 &      26.391 & 0.22 & 0.79 &     3 &   409 &   E  \\
  394 & Arduina$^\ast$            &     0 & $-79$ &   191 & $-49$ & 16.62157 &        16.5 & 0.28 & 0.54 &     3 &   395 &  CE  \\
  513 & Centesima                 &   149 &     4 &   332 &    15 &  5.32399 &        5.23 & 0.18 & 0.45 &     3 &   460 &   C  \\
  636 & Erika                     &    13 & $-70$ &   176 & $-60$ & 14.60771 &      14.603 & 0.29 & 0.33 &     3 &   498 &   E  \\
  670 & Ottegebe$^\ast$           &   127 &    77 &   304 &    68 & 10.03991 &      10.045 & 0.34 & 0.35 &     3 &   540 &   C  \\
  682 & Hagar$^\ast$              &    93 & $-71$ &   277 & $-35$ & 4.850417 &      4.8503 & 0.49 & 0.52 &     3 &   334 &   C  \\
  700 & Auravictrix               &    54 &    33 &   249 &    47 &  6.07489 &       6.075 & 0.18 & 0.43 &     3 &   404 &  CE  \\
  706 & Hirundo$^\ast$            &    91 &    70 &   250 &    45 &  22.0161 &      22.027 & 0.39 &  0.9 &     3 &   365 &   E  \\
  744 & Aguntina                  &    44 & $-58$ &   227 & $-51$ &  17.4690 &       17.47 &      & 0.50 &     3 &   388 &  CE  \\
  762 & Pulcova$^\ast$            &    20 & $-12$ &   196 & $-42$ &  5.83977 &       5.839 & 0.18 & 0.30 &     3 &   408 &   E  \\
  769 & Tatjana                   &   176 &    54 &   347 &    38 &  35.0637 &       35.08 & 0.30 & 0.33 &  3$-$ &   437 &   E  \\
  844 & Leontina                  &   302 &    68 &       &       &  6.78303 &      6.7859 & 0.20 & 0.26 &     3 &   415 &   E  \\
  885 & Ulrike                    &    13 & $-64$ &   207 & $-60$ & 4.906164 &        4.90 & 0.55 & 0.72 &     3 &   416 &   C  \\
  918 & Itha                      &    59 & $-59$ &   249 & $-72$ & 3.473810 &     3.47393 & 0.15 & 0.30 &     3 &   350 &   E  \\
  943 & Begonia                   &   209 & $-75$ &       &       &  15.6593 &       15.66 & 0.24 & 0.34 &     3 &   363 &   E  \\
  968 & Petunia                   &   355 & $-78$ &       &       &   61.160 &      61.280 &      & 0.38 &     3 &   361 &  CE  \\
  979 & Ilsewa                    &   352 & $-66$ &       &       &  42.8982 &       42.61 & 0.30 & 0.31 &     3 &   370 &   E  \\
 1077 & Campanula                 &   178 &    76 &   313 &    59 & 3.850486 &     3.85085 & 0.24 & 0.40 &     3 &   361 &   E  \\
 1083 & Salvia                    &   165 & $-59$ &   358 & $-58$ & 4.281429 &        4.23 &      & 0.61 &     3 &   277 &   C  \\
 1150 & Achaia$^\ast$             &   126 & $-65$ &   315 & $-60$ &   61.071 &       60.99 &      & 0.72 &     3 &   396 &  CE  \\
 1294 & Antwerpia                 &   128 & $-66$ &   246 & $-76$ &  6.62521 &        6.63 & 0.35 & 0.42 &     3 &   317 &   E  \\
 1332 & Marconia                  &    37 &    31 &   220 &    31 &  19.2264 &       19.16 &      & 0.30 &     3 &   408 &   E  \\
 1352 & Wawel$^\ast$              &    37 &    55 &   201 &    52 &  16.9542 &       16.97 & 0.35 & 0.44 &     3 &   356 &   C  \\
 1379 & Lomonosowa                &    72 & $-84$ &   265 & $-46$ &  24.4846 &      24.488 &      & 0.63 &     3 &   359 &   E  \\
 1407 & Lindelof                  &   147 &    36 &       &       &  31.0941 &      31.151 &      & 0.34 &     3 &   397 &  CE  \\
 1449 & Virtanen$^\ast$           &    89 &    61 &   302 &    61 &  30.5006 &      30.495 & 0.08 & 0.69 &  3$-$ &   354 &   E  \\
 1486 & Marilyn$^\ast$            &    83 & $-57$ &   270 & $-62$ & 4.566945 &       4.566 & 0.40 & 0.48 &     3 &   492 &   C  \\
 1496 & Turku                     &    75 & $-75$ &       &       &  6.47375 &        6.47 &      & 0.51 &  3$-$ &   486 &  CE  \\
 1521 & Seinajoki                 &    63 & $-18$ &   230 & $-37$ & 4.328159 &        4.32 &      & 0.15 &     3 &   331 &   C  \\
 1592 & Mathieu                   &    94 & $-14$ &   269 & $ -7$ &  28.4821 &       28.46 &      & 0.50 &    2+ &   302 &   E  \\
 1716 & Peter                     &    52 & $-60$ &   252 & $-52$ & 11.51720 &      11.514 &      & 0.52 &     3 &   348 &  CE  \\
 1723 & Klemola                   &    52 & $-54$ &   239 & $-56$ &  6.25609 &      6.2545 & 0.16 & 0.33 &     3 &   484 &   E  \\
 1820 & Lohmann$^\ast$            &    62 &    67 &   250 &    58 & 14.04497 &     14.0554 & 0.40 & 0.55 &     3 &   281 &   E  \\
 1833 & Shmakova                  &    88 &    47 &   336 &    84 & 3.838235 &        3.93 &      & 0.38 &     3 &   355 &   E  \\
 1858 & Lobachevskij              &    80 &    50 &   255 &    48 &  5.41208 &       5.413 & 0.30 & 0.48 &    2+ &   390 &   E  \\
 1860 & Barbarossa                &    45 &    30 &   238 &    63 & 3.254853 &       3.255 & 0.28 & 0.35 &     3 &   404 &   E  \\
 1906 & Naef                      &    72 & $-70$ &       &       & 11.00818 &      11.009 & 0.92 & 0.96 &     3 &   319 &   C  \\
 2064 & Thomsen                   &   118 & $-72$ &   334 & $-56$ & 4.244023 &       4.233 &      & 0.62 &     3 &   550 &   E  \\
 2275 & Cuitlahuac                &     9 & $-65$ &       &       &  6.29005 &      6.2891 &      & 1.10 &     3 &   536 &   E  \\
 3015 & Candy                     &   142 & $-26$ &   346 & $-70$ & 4.625223 &       4.625 & 0.50 & 1.05 &     3 &   450 &   E  \\
 3773 & Smithsonian$^\ast$        &   121 & $-62$ &       &       &  6.98132 &      6.9804 &      & 1.04 &     3 &   622 &   C  \\
 4089 & Galbraith                 &    64 &    69 &   218 &    50 &  4.91316 &      4.9123 &      & 0.68 &     3 &   687 &   C  \\
 4265 & Kani$^\ast$               &    74 &    67 &   277 &    72 & 5.727574 &      5.7279 &      & 0.75 &     3 &   730 &  CE  \\
 4641 & 1990 QT3                  &   178 & $-46$ &   359 & $-50$ & 5.313079 &      5.3126 &      & 0.88 &     3 &   541 &   C  \\
 4801 & Ohre                      &   121 & $-74$ &   266 & $-85$ &  31.9990 &      32.000 & 0.50 & 0.60 &     3 &   581 &  CE  \\
 4896 & Tomoegozen                &   276 & $-66$ &       &       &  8.87996 &       8.869 &      & 0.65 &     3 &   371 &  CE  \\
 4995 & 1984 QR                   &   243 &    68 &   355 &    70 &  26.3920 &       26.37 &      & 0.82 &  3$-$ &   261 &   E  \\
 5738 & Billpickering             &     2 & $-66$ &       &       & 10.38264 &        10.4 &      & 0.45 &     3 &   118 &   E  \\
 6406 & 1992 MJ$^\ast$            &    17 & $-61$ &   216 & $-52$ &  6.81816 &       6.819 & 1.10 & 1.18 &     3 &   508 &  CE  \\
 6487 & Tonyspear                 &   165 & $-90$ &       &       &   74.501 &       74.91 &      & 1.24 &  3$-$ &   372 &   C  \\
 6510 & Tarry                     &    84 & $-72$ &   249 & $-37$ &  6.36490 &       6.370 & 0.50 & 0.54 &     3 &   376 &   E  \\
 7783 & 1994 JD                   &    70 & $-71$ &       &       &  31.8661 &       31.83 &      & 0.85 &     3 &   265 &   E  \\
 9983 & Rickfienberg              &   120 & $-58$ &   248 & $-77$ &  5.29616 &      5.2963 &      & 1.30 &     3 &   371 &   C  \\
12045 & Klein                     &    77 & $-22$ &       &       &  9.00648 &      8.9686 &      & 0.55 &  3$-$ &   453 &   E  \\
14257 & 2000 AR97                 &    57 &    67 &       &       & 13.57929 &      13.584 &      & 0.67 &     3 &   514 &   C  \\
28887 & 2000 KQ58$^\ast$          &    11 & $-61$ &   191 & $-21$ &  6.84315 &      6.8429 &      & 0.55 &     3 &   368 &   E  \\
31485 & 1999 CM51                 &    94 & $-36$ &   316 & $-56$ &  6.00262 &       6.001 & 0.65 & 0.68 &     3 &   375 &   E  \\
34484 & 2000 SR124$^\ast$         &   236 & $-55$ &       &       &  6.17519 &       6.174 &      & 0.80 &    2+ &   473 &   C  \\
44600 & 1999 RU10                 &   267 & $-24$ &       &       &  6.21130 &       6.211 & 0.98 & 1.09 &     3 &   310 &   C  \\
80276 & 1999 XL32                 &    60 &    38 &   275 &    63 &  5.57148 &        5.56 &      & 0.77 &  3$-$ &   258 &   C  \\
88161 & 2000 XK18                 &   195 &    53 &   340 &    70 &  6.80480 &       6.806 &      & 0.80 &     3 &   429 &   C  \\
    \end{longtable}
  \end{longtab}

  \begin{acknowledgements}
    We would not be able to process data for hundreds of thousands of asteroids without the help of tens of thousand of volunteers who joined the Asteroids@home BOINC project and provided computing resources from their computers. We greatly appreciate their contribution. The work of J\v{D} was supported by the grant 15-04816S of the Czech Science Foundation. JH greatly appreciates the CNES post-doctoral fellowship program. JH was supported by the project under the contract 11-BS56-008 (SHOCKS) of the French Agence National de la Recherche (ANR). DO was supported by the grant NCN 2012/S/ST9/00022 Polish National Science Center. We thank the referee A.~Marciniak for providing constructive comments that improved the contents of this paper.
  \end{acknowledgements}

\newcommand{\SortNoop}[1]{}


\begin{thebibliography}{37}
\expandafter\ifx\csname natexlab\endcsname\relax\def\natexlab#1{#1}\fi

\bibitem[{{Bowell} {et~al.}(2014){Bowell}, {Oszkiewicz}, {Wasserman},
  {Muinonen}, {Penttil{\"a}}, \& {Trilling}}]{Bow.ea:14}
{Bowell}, E., {Oszkiewicz}, D.~A., {Wasserman}, L.~H., {et~al.} 2014,
  Meteoritics and Planetary Science, 49, 95

\bibitem[{{Cellino} {et~al.}(2009){Cellino}, {Hestroffer}, {Tanga}, {Mottola},
  \& {Dell'Oro}}]{Cel.ea:09}
{Cellino}, A., {Hestroffer}, D., {Tanga}, P., {Mottola}, S., \& {Dell'Oro}, A.
  2009, \aap, 506, 935

\bibitem[{{Clark}(2014)}]{Cla:14}
{Clark}, M. 2014, Minor Planet Bulletin, 41, 178

\bibitem[{{\SortNoop{Durech05}{\v D}urech}
  {et~al.}(2005){\SortNoop{Durech05}{\v D}urech}, {Grav}, {Jedicke},
  {Kaasalainen}, \& {Denneau}}]{Dur.ea:05}
{\SortNoop{Durech05}{\v D}urech}, J., {Grav}, T., {Jedicke}, R., {Kaasalainen},
  M., \& {Denneau}, L. 2005, Earth, Moon, and Planets, 97, 179

\bibitem[{{\SortNoop{Durech06}\v{D}urech}
  {et~al.}(2007){\SortNoop{Durech06}\v{D}urech}, {Scheirich}, {Kaasalainen},
  {Grav}, {Jedicke}, \& {Denneau}}]{Dur.ea:07}
{\SortNoop{Durech06}\v{D}urech}, J., {Scheirich}, P., {Kaasalainen}, M.,
  {et~al.} 2007, in {Near Earth Objects, our Celestial Neighbors: Opportunity
  and Risk}, ed. A.~{Milani}, G.~B. {Valsecchi}, \& D.~{Vokrouhlick\'y}
  (Cambridge: {Cambridge University Press}), 191

\bibitem[{{\SortNoop{Durech10}{\v D}urech}
  {et~al.}(2009){\SortNoop{Durech10}{\v D}urech}, {Kaasalainen}, {Warner},
  {Fauerbach}, {Marks}, {Fauvaud}, {Fauvaud}, {Vugnon}, {Pilcher},
  {Bernasconi}, \& {Behrend}}]{Dur.ea:09}
{\SortNoop{Durech10}{\v D}urech}, J., {Kaasalainen}, M., {Warner}, B.~D.,
  {et~al.} 2009, \aap, 493, 291

\bibitem[{{\SortNoop{Durech11}\v{D}urech}
  {et~al.}(2010){\SortNoop{Durech11}\v{D}urech}, {Sidorin}, \&
  {Kaasalainen}}]{Dur.ea:10}
{\SortNoop{Durech11}\v{D}urech}, J., {Sidorin}, V., \& {Kaasalainen}, M. 2010,
  \aap, 513, A46

\bibitem[{{\SortNoop{Durech15}\v{D}urech}
  {et~al.}(2016){\SortNoop{Durech15}\v{D}urech}, {Carry}, {Delbo},
  {Kaasalainen}, \& {Viikinkoski}}]{Dur.ea:15b}
{\SortNoop{Durech15}\v{D}urech}, J., {Carry}, B., {Delbo}, M., {Kaasalainen},
  M., \& {Viikinkoski}, M. 2016, in {Asteroids~IV, in press}, ed. P.~{Michel},
  F.~{DeMeo}, \& W.~{Bottke} (Tucson: {University of Arizona Press})

\bibitem[{{\SortNoop{Durech15}{\v D}urech}
  {et~al.}(2015){\SortNoop{Durech15}{\v D}urech}, {Hanu{\v s}}, \& {Van{\v
  c}o}}]{Dur.ea:15}
{\SortNoop{Durech15}{\v D}urech}, J., {Hanu{\v s}}, J., \& {Van{\v c}o}, R.
  2015, Astronomy and Computing, 13, 80

\bibitem[{{\SortNoop{Durech16}\v{D}urech}
  {et~al.}(2016){\SortNoop{Durech16}\v{D}urech}, {Hanu\v{s}}, {Ali-Lagoa},
  {Delbo}, \& {Oszkiewicz}}]{Dur.ea:15c}
{\SortNoop{Durech16}\v{D}urech}, J., {Hanu\v{s}}, J., {Ali-Lagoa}, V., {Delbo},
  M., \& {Oszkiewicz}, D. 2016, in {Proceedings IAU Symposium No. 318, in
  press}, ed. S.~{Chesley}, R.~{Jedicke}, A.~{Morbidelli}, \& D.~{Farnocchia}
  (Cambridge: {Cambridge University Press})

\bibitem[{{Gary}(2004)}]{Gar:04}
{Gary}, B.~L. 2004, Minor Planet Bulletin, 31, 56

\bibitem[{{Hanu{\v s}} {et~al.}(2013{\natexlab{a}}){Hanu{\v s}}, {Bro{\v z}},
  {Durech}, {Warner}, {Brinsfield}, {Durkee}, {Higgins}, {Koff}, {Oey},
  {Pilcher}, {Stephens}, {Strabla}, {Ulisse}, \& {Girelli}}]{Han.ea:13c}
{Hanu{\v s}}, J., {Bro{\v z}}, M., {Durech}, J., {et~al.} 2013{\natexlab{a}},
  \aap, 559, A134

\bibitem[{{Hanu{\v s}} {et~al.}(2013{\natexlab{b}}){Hanu{\v s}}, {Marchis}, \&
  {{\v D}urech}}]{Han.ea:13a}
{Hanu{\v s}}, J., {Marchis}, F., \& {{\v D}urech}, J. 2013{\natexlab{b}},
  \icarus, 226, 1045

\bibitem[{{Hanu{\v s}} {et~al.}(2013{\natexlab{c}}){Hanu{\v s}}, {{\v D}urech},
  {Bro{\v z}}, {Marciniak}, {Warner}, {Pilcher}, {Stephens}, {Behrend},
  {Carry}, {{\v C}apek}, {Antonini}, {Audejean}, {Augustesen}, {Barbotin},
  {Baudouin}, {Bayol}, {Bernasconi}, {Borczyk}, {Bosch}, {Brochard},
  {Brunetto}, {Casulli}, {Cazenave}, {Charbonnel}, {Christophe}, {Colas},
  {Coloma}, {Conjat}, {Cooney}, {Correira}, {Cotrez}, {Coupier}, {Crippa},
  {Cristofanelli}, {Dalmas}, {Danavaro}, {Demeautis}, {Droege}, {Durkee},
  {Esseiva}, {Esteban}, {Fagas}, {Farroni}, {Fauvaud}, {Fauvaud}, {Del Freo},
  {Garcia}, {Geier}, {Godon}, {Grangeon}, {Hamanowa}, {Hamanowa}, {Heck},
  {Hellmich}, {Higgins}, {Hirsch}, {Husarik}, {Itkonen}, {Jade},
  {Kami{\'n}ski}, {Kankiewicz}, {Klotz}, {Koff}, {Kryszczy{\'n}ska},
  {Kwiatkowski}, {Laffont}, {Leroy}, {Lecacheux}, {Leonie}, {Leyrat},
  {Manzini}, {Martin}, {Masi}, {Matter}, {Micha{\l}owski}, {Micha{\l}owski},
  {Micha{\l}owski}, {Michelet}, {Michelsen}, {Morelle}, {Mottola}, {Naves},
  {Nomen}, {Oey}, {Og{\l}oza}, {Oksanen}, {Oszkiewicz},
  {P{\"a}{\"a}kk{\"o}nen}, {Paiella}, {Pallares}, {Paulo}, {Pavic}, {Payet},
  {Poli{\'n}ska}, {Polishook}, {Poncy}, {Revaz}, {Rinner}, {Rocca}, {Roche},
  {Romeuf}, {Roy}, {Saguin}, {Salom}, {Sanchez}, {Santacana}, {Santana-Ros},
  {Sareyan}, {Sobkowiak}, {Sposetti}, {Starkey}, {Stoss}, {Strajnic}, {Teng},
  {Tr{\'e}gon}, {Vagnozzi}, {Velichko}, {Waelchli}, {Wagrez}, \&
  {W{\"u}cher}}]{Han.ea:13b}
{Hanu{\v s}}, J., {{\v D}urech}, J., {Bro{\v z}}, M., {et~al.}
  2013{\natexlab{c}}, \aap, 551, A67

\bibitem[{{Hanu{\v s}} {et~al.}(2011){Hanu{\v s}}, {{\v D}urech}, {Bro{\v z}},
  {Warner}, {Pilcher}, {Stephens}, {Oey}, {Bernasconi}, {Casulli}, {Behrend},
  {Polishook}, {Henych}, {Lehk{\'y}}, {Yoshida}, \& {Ito}}]{Han.ea:11}
{Hanu{\v s}}, J., {{\v D}urech}, J., {Bro{\v z}}, M., {et~al.} 2011, \aap, 530,
  A134

\bibitem[{{Hanu{\v s}} {et~al.}(2016){Hanu{\v s}}, {{\v D}urech}, {Oszkiewicz},
  {Behrend}, {Carry}, {Delbo'}, {Adam}, {Afonina}, {Anquetin}, {Antonini},
  {Arnold}, {Audejean}, {Aurard}, {Bachschmidt}, {Badue}, {Barbotin}, {Barroy},
  {Baudouin}, {Berard}, {Berger}, {Bernasconi}, {Bosch}, {Bouley}, {Bozhinova},
  {Brinsfield}, {Brunetto}, {Canaud}, {Caron}, {Carrier}, {Casalnuovo},
  {Casulli}, {Cerda}, {Chalamet}, {Charbonnel}, {Chinaglia}, {Cikota}, {Colas},
  {Coliac}, {Collet}, {Coloma}, {Conjat}, {Conseil}, {Costa}, {Crippa},
  {Cristofanelli}, {Damerdji}, {Debackere}, {Decock}, {D{\'e}hais},
  {D{\'e}l{\'e}age}, {Delmelle}, {Demeautis}, {Dr{\'o}{\.z}d{\.z}}, {Dubos},
  {Dulcamara}, {Dumont}, {Durkee}, {Dymock}, {Escalante del Valle}, {Esseiva},
  {Esseiva}, {Esteban}, {Fauchez}, {Fauerbach}, {Fauvaud}, {Fauvaud},
  {Forn{\'e}}, {Fournel}, {Fradet}, {Garlitz}, {Gerteis}, {Gillier}, {Gillon},
  {Giraud}, {Godard}, {Goncalves}, {Hamanowa}, {Hamanowa}, {Hay}, {Hellmich},
  {Heterier}, {Higgins}, {Hirsch}, {Hodosan}, {Hren}, {Hygate}, {Innocent},
  {Jacquinot}, {Jawahar}, {Jehin}, {Jerosimic}, {Klotz}, {Koff}, {Korlevic},
  {Kosturkiewicz}, {Krafft}, {Krugly}, {Kugel}, {Labrevoir}, {Lecacheux},
  {Lehk{\'y}}, {Leroy}, {Lesquerbault}, {Lopez-Gonzales}, {Lutz}, {Mallecot},
  {Manfroid}, {Manzini}, {Marciniak}, {Martin}, {Modave}, {Montaigut},
  {Montier}, {Morelle}, {Morton}, {Mottola}, {Naves}, {Nomen}, {Oey},
  {Og{\l}oza}, {Paiella}, {Pallares}, {Peyrot}, {Pilcher}, {Pirenne}, {Piron},
  {Polinska}, {Polotto}, {Poncy}, {Previt}, {Reignier}, {Renauld}, {Ricci},
  {Richard}, {Rinner}, {Risoldi}, {Robilliard}, {Romeuf}, {Rousseau}, {Roy},
  {Ruthroff}, {Salom}, {Salvador}, {Sanchez}, {Santana-Ros}, {Scholz},
  {S{\'e}n{\'e}}, {Skiff}, {Sobkowiak}, {Sogorb}, {Sold{\'a}n}, {Spiridakis},
  {Splanska}, {Sposetti}, {Starkey}, {Stephens}, {Stiepen}, {Stoss},
  {Strajnic}, {Teng}, {Tumolo}, {Vagnozzi}, {Vanoutryve}, {Vugnon}, {Warner},
  {Waucomont}, {Wertz}, {Winiarski}, \& {Wolf}}]{Han.ea:15b}
{Hanu{\v s}}, J., {{\v D}urech}, J., {Oszkiewicz}, D.~A., {et~al.} 2016, \aap,
  in press

\bibitem[{{Hawkins} \& {Ditteon}(2008)}]{Haw.Dit:08}
{Hawkins}, S. \& {Ditteon}, R. 2008, Minor Planet Bulletin, 35, 1

\bibitem[{{Johansen} \& {Lacerda}(2010)}]{Joh.Lac:10}
{Johansen}, A. \& {Lacerda}, P. 2010, \mnras, 404, 475

\bibitem[{{Kaasalainen}(2004)}]{Kaa:04}
{Kaasalainen}, M. 2004, \aap, 422, L39

\bibitem[{{Kaasalainen} \& {\v{D}urech}(2007)}]{Kaa.Dur:07}
{Kaasalainen}, M. \& {\v{D}urech}, J. 2007, in {Near Earth Objects, our
  Celestial Neighbors: Opportunity and Risk}, ed. A.~{Milani}, G.~B.
  {Valsecchi}, \& D.~{Vokrouhlick\'y} (Cambridge: {Cambridge University
  Press}), 151

\bibitem[{{Kaasalainen} \& {Lamberg}(2006)}]{Kaa.Lam:06}
{Kaasalainen}, M. \& {Lamberg}, L. 2006, Inverse Problems, 22, 749

\bibitem[{{Kaasalainen} {et~al.}(2002){Kaasalainen}, {Mottola}, \&
  {Fulchignomi}}]{Kaa.ea:02c}
{Kaasalainen}, M., {Mottola}, S., \& {Fulchignomi}, M. 2002, in
  {Asteroids~III}, ed. W.~F. {Bottke}, A.~{Cellino}, P.~{Paolicchi}, \& R.~P.
  {Binzel} (Tucson: {University of Arizona Press}), 139--150

\bibitem[{{Kaasalainen} \& {Torppa}(2001)}]{Kaa.Tor:01}
{Kaasalainen}, M. \& {Torppa}, J. 2001, Icarus, 153, 24

\bibitem[{{Kaasalainen} {et~al.}(2001){Kaasalainen}, {Torppa}, \&
  {Muinonen}}]{Kaa.ea:01}
{Kaasalainen}, M., {Torppa}, J., \& {Muinonen}, K. 2001, Icarus, 153, 37

\bibitem[{{Kryszczy{\'n}ska} {et~al.}(2007){Kryszczy{\'n}ska}, {La Spina},
  {Paolicchi}, {Harris}, {Breiter}, \& {Pravec}}]{Kry.ea:07}
{Kryszczy{\'n}ska}, A., {La Spina}, A., {Paolicchi}, P., {et~al.} 2007, Icarus,
  192, 223

\bibitem[{{La Spina} {et~al.}(2004){La Spina}, {Paolicchi}, {Kryszczy{\'n}ska},
  \& {Pravec}}]{LaS.ea:04}
{La Spina}, A., {Paolicchi}, P., {Kryszczy{\'n}ska}, A., \& {Pravec}, P. 2004,
  \nat, 428, 400

\bibitem[{{Lagerkvist}(1978)}]{Lag:78}
{Lagerkvist}, C.-I. 1978, \aaps, 31, 361

\bibitem[{{Marciniak} {et~al.}(2015){Marciniak}, {Pilcher}, {Oszkiewicz},
  {Santana-Ros}, {Urakawa}, {Fauvaud}, {Kankiewicz}, {Tychoniec}, {Fauvaud},
  {Hirsch}, {Horbowicz}, {Kami{\'n}ski}, {Konstanciak}, {Kosturkiewicz},
  {Murawiecka}, {Nadolny}, {Nishiyama}, {Okumura}, {Poli{\'n}ska}, {Richard},
  {Sakamoto}, {Sobkowiak}, {Stachowski}, \& {Trela}}]{Mar.ea:15}
{Marciniak}, A., {Pilcher}, F., {Oszkiewicz}, D., {et~al.} 2015, \planss, 118,
  256

\bibitem[{{Ostro} \& {Connelly}(1984)}]{Ost.Con:84}
{Ostro}, S.~J. \& {Connelly}, R. 1984, Icarus, 57, 443

\bibitem[{Oszkiewicz {et~al.}(2011)Oszkiewicz, Muinonen, Bowell, Trilling,
  Penttil{\"a}, Pieniluoma, Wasserman, \& Enga}]{Osz.ea:11}
Oszkiewicz, D., Muinonen, K., Bowell, E., {et~al.} 2011, Journal of
  Quantitative Spectroscopy and Radiative Transfer, 112, 1919, electromagnetic
  and Light Scattering by Nonspherical Particles \{XII\}

\bibitem[{{Pravec} {et~al.}(2002){Pravec}, {Harris}, \&
  {Micha{\l}owski}}]{Pra.ea:02b}
{Pravec}, P., {Harris}, A.~W., \& {Micha{\l}owski}, T. 2002, in
  {Asteroids~III}, ed. W.~F. {Bottke}, A.~{Cellino}, P.~{Paolicchi}, \& R.~P.
  {Binzel} (Tucson: {University of Arizona Press}), 113--122

\bibitem[{{Santana-Ros} {et~al.}(2015){Santana-Ros}, {Bartczak},
  {Micha{\l}owski}, {Tanga}, \& {Cellino}}]{San.ea:15}
{Santana-Ros}, T., {Bartczak}, P., {Micha{\l}owski}, T., {Tanga}, P., \&
  {Cellino}, A. 2015, \mnras, 450, 333

\bibitem[{{Torppa} {et~al.}(2003){Torppa}, {Kaasalainen}, {Michalowski},
  {Kwiatkowski}, {Kryszczy{\' n}ska}, {Denchev}, \& {Kowalski}}]{Tor.ea:03}
{Torppa}, J., {Kaasalainen}, M., {Michalowski}, T., {et~al.} 2003, Icarus, 164,
  346

\bibitem[{{Warner} {et~al.}(2009){Warner}, {Harris}, \& {Pravec}}]{War.ea:09}
{Warner}, B.~D., {Harris}, A.~W., \& {Pravec}, P. 2009, Icarus, 202, 134

\bibitem[{{Waszczak} {et~al.}(2015){Waszczak}, {Chang}, {Ofek}, {Laher},
  {Masci}, {Levitan}, {Surace}, {Cheng}, {Ip}, {Kinoshita}, {Helou}, {Prince},
  \& {Kulkarni}}]{Was.ea:15}
{Waszczak}, A., {Chang}, C.-K., {Ofek}, E.~O., {et~al.} 2015, \aj, 150, 75

\bibitem[{{Wisniewski} {et~al.}(1997){Wisniewski}, {Michalowski}, {Harris}, \&
  {McMillan}}]{Wis.ea:97}
{Wisniewski}, W.~Z., {Michalowski}, T.~M., {Harris}, A.~W., \& {McMillan},
  R.~S. 1997, Icarus, 126, 395

\bibitem[{{Wright} {et~al.}(2010){Wright}, {Eisenhardt}, {Mainzer}, {Ressler},
  {Cutri}, {Jarrett}, {Kirkpatrick}, {Padgett}, {McMillan}, {Skrutskie},
  {Stanford}, {Cohen}, {Walker}, {Mather}, {Leisawitz}, {Gautier}, {McLean},
  {Benford}, {Lonsdale}, {Blain}, {Mendez}, {Irace}, {Duval}, {Liu}, {Royer},
  {Heinrichsen}, {Howard}, {Shannon}, {Kendall}, {Walsh}, {Larsen}, {Cardon},
  {Schick}, {Schwalm}, {Abid}, {Fabinsky}, {Naes}, \& {Tsai}}]{Wri.ea:10}
{Wright}, E.~L., {Eisenhardt}, P.~R.~M., {Mainzer}, A.~K., {et~al.} 2010, \aj,
  140, 1868

\end{thebibliography}
\end{document}